\documentclass[aps,prl,superscriptaddress,twocolumn]{revtex4-2}

\usepackage{comment}
\usepackage{graphicx}
\usepackage{dcolumn}
\usepackage{bm}
\usepackage{physics}
\usepackage{color}
\usepackage[dvipsnames]{xcolor}
\usepackage{url}
\usepackage{hyperref}
\hypersetup{%
   pdfpagemode=UseNone, 
   pdfstartpage=1,
   pdfmenubar=true,
   pdftoolbar=true,
   colorlinks = true,
   linkcolor=blue,
   citecolor=blue,
   urlcolor=blue,
   bookmarksopen=false
 }
\usepackage[normalem]{ulem}
\usepackage{bbold}

\begin{document}


\title{Generalized Gross-Pitaevskii Equation for 2D Bosons with Attractive Interactions
}%

\author{Michał Suchorowski}
 \email{m.suchorowski@uw.edu.pl}
\affiliation{%
  Faculty of Physics, University of Warsaw, Pasteura 5, 02-093 Warsaw, Poland
}%

\author{Fabian Brauneis}
\affiliation{Department of Physics, Technische Universit{\"a}t Darmstadt, 64289 Darmstadt, Germany}%
\author{Hans-Werner Hammer}
\affiliation{Department of Physics, Technische Universit{\"a}t Darmstadt, 64289 Darmstadt, Germany} \affiliation{Extreme Matter Institute EMMI and Helmholtz Forschungsakademie Hessen f{\"u}r FAIR (HFHF),
GSI Helmholtzzentrum f{\"u}r Schwerionenforschung GmbH, 64291 Darmstadt, Germany
}%
\author{Michał Tomza}
\affiliation{Faculty of Physics, University of Warsaw, Pasteura 5, 02-093 Warsaw, Poland
}%
\author{Artem G. Volosniev}
\email{artem@phys.au.dk }
\affiliation{Center for Complex Quantum Systems, Department of Physics and Astronomy, Aarhus University, Ny Munkegade 120, DK-8000 Aarhus C, Denmark
}%

\date{\today}

\begin{abstract}
 {\color{black}  We introduce a generalized Gross-Pitaevskii equation that provides a nonlinear framework for studying two-dimensional (2D) attractive Bose systems. Its defining feature is the logarithmic density dependence of the coupling constant, which breaks the scale invariance inherent in the standard mean-field equations. This framework allows straightforward calculations of the system properties arising from the quantum anomaly.  As a first illustration, we study universal bound states in free space, commonly referred to as quantum droplets. Then, we analyze breathing modes and quench dynamics in trapped systems, paving the way for a systematic exploration of non-equilibrium phenomena in 2D attractive Bose systems. Finally, we predict the existence of universal excited states, including vortex configurations, which may be more accessible to experimental investigation than the ground state. Our results provide a robust theoretical foundation for studying both static and dynamical properties of finite systems, and offer guidance for the design of future experiments}.
\end{abstract}

\maketitle


Scale invariance of the attractive two-dimensional (2D) Gross-Pitaevskii equation (GPE) gives rise to unique phenomena such as Townes solitons~\cite{Chiao1964}, `strong' self-similar collapse~\cite{strongcollapse,strongcollapse1}, and interaction-independent breathing-mode frequencies in tightly trapped systems~\cite{Pitaevskii1996, Pitaevskii1997}.  
\textcolor{black}{
In realistic quantum systems of single-component cold bosons with short-range interactions, this symmetry is, however, broken by the presence of the two-body bound state that introduces an explicit length scale into the problem. One of the most striking consequences of this `quantum anomaly'~\cite{Jackiw:1991je,Olshanii2010} is the formation of universal many-body bound states~\cite{HammerPRL2004,rajeev1999a,rajeev1999b}.
For a finite number of particles, the balance between kinetic energy and short-range interactions confines these states to a finite size, thus justifying their designation as quantum droplets. While their microscopic nature is clearly distinct from that of quantum droplets in two-component or dipolar Bose gases~\cite{Luo2020,Chomaz2022,Mistakidis2023}, they likewise arise from quantum fluctuations.
}
 
Despite this exciting progress in our understanding of 2D attractive bosons, it remains unclear whether a simple, unified theoretical framework exists to analyze these phenomena.   Consequently, each new solution represents a state-of-the-art effort; see, for example, recent studies in Refs.~\cite{Petrov2024, Tononi2024, Brauneis2024}, {\color{black} which rely on beyond-mean-field, quantum Monte Carlo and variational approaches, respectively. 
 Here, to address this issue, we introduce a generalized GPE that incorporates a physical length scale via a density-dependent coupling, see `Framework' below. This equation enables a straightforward and unified description of both the static and dynamic properties of the system, including excited droplet states.

To illustrate the framework, we first demonstrate that it successfully} describes the behavior of universal droplets~\cite{HammerPRL2004,Petrov2024}, while preserving a simple structure suitable for intuitive analytical and numerical exploration. {\color{black} For example, in our theory, the finite droplet size emerges naturally because the coupling constant vanishes at high densities, thereby preventing collapse due to attractive interactions (see `Benchmark' below). 
We note in passing that, because the coupling vanishes in the high-density limit, 2D bosons may become a useful toy model for the renormalization of strong interactions in particle physics~\cite{rajeev1999b}. Our methods and results may therefore be useful for building intuition about Quantum Chromodynamics (QCD) systems, e.g., quark matter with vortices~\cite{Iida:2002ev,Eto:2009tr}.} Indeed, QCD is asymptotically free~\cite{Gross:1973id,Politzer:1973fx} and thus naively scale invariant at high energies, but this scale invariance is anomalous and broken by dimensional transmutation.
In finite density QCD, a density-dependent coupling constant enters naturally~\cite{Collins:1974ky}, although in practice, the chemical potential is often used instead of the density~\cite{Kurkela:2009gj}.

{\color{black} In `Discussion', we demonstrate the capability of the generalized GPE by examining the universal ground state in a trap, and discussing corresponding breathing dynamics. Finally, we summarize our findings and outline prospects for future studies of quench dynamics and universal excited states, such as vortices.}

{\color{black}
{\it Framework: Energy Functional.} 
We study an ultradilute system of $N$ 2D bosons of mass $m$.
In our analysis focused on the limit $N\gg 1$, we rely on the following approximation for the energy }
\begin{equation}
\frac{E_N[\psi]}{N}=\int \mathrm{d}\mathbf{x} \psi^*\left[-\frac{\hbar^2}{2m} \pdv[2]{\mathbf{x}}  +W(\mathbf{x})+\frac{g N}{2}|\psi|^2\right]\psi,
\label{eq:energy}
\end{equation}
where $\psi=\psi(\mathbf{x})$ is a normalized function ($\norm{\psi}=1$) found by minimization of $E_N$; $W$ is the trapping potential.  To facilitate a clear and consistent illustration of our results, we either consider a free system ($W=0$) or employ a harmonic trap, $W(\mathbf{x})=m \omega^2 \mathbf{x}^2/2$.
In what follows, we use the units with $\hbar=m=\omega=1$. Energy is then expressed in the units of $\hbar \omega$; interaction strength, $g$, in $\hbar^2/m$; length in $\sqrt{\hbar/(m\omega)}$; time and frequencies in $1/\omega$ and $\omega$, respectively.  We will also often express length in the units of the state size, $R=\sqrt{\int\mathrm{d}\mathbf{x} |\mathbf{x} \psi|^2}$, which is the only relevant length scale for $W=0$ or equivalently in the regime of strong attractions.

A unique aspect of 2D many-boson systems is that their theoretical description requires the coupling constant $g$ to depend on the system's probability density, $n=|\psi|^2$, even in the low-energy (i.e., weakly-interacting) regime.  For repulsive boson-boson interactions, $g>0$, a widely used expression is $g=4\pi / |\ln(N na^2)|$~\cite{Schick1971,Shevchenko1992,LiebCMP2001}, where $a$ denotes the 2D scattering length.
The logarithmic dependence of $g$ on $n$ is notably weak, which poses challenges for experimental detection~\cite{Kolomeisky2000}. Nevertheless, this density dependence has a fundamental impact -- it underlies the aforementioned symmetry breaking, as the scale invariance of the mean-field approximation relies on the assumption that $g$ remains constant.

 For attractive interactions and in the absence of an external potential ($W=0$), a possible form of the coupling constant is 
$g\simeq -4\pi/|\ln(R^2 B_2)|$~\cite{HammerPRL2004}, where $B_2>0$ denotes the two-body binding energy. This expression is, however, not as practically useful as the one for repulsive interactions. In particular, it cannot be straightforwardly extended to include an external trap -- essential for modeling cold-atom systems -- since the parameter $R$ must then be determined from a separate equation, which is challenging to solve numerically~\cite{Brauneis2024}. Furthermore, it is unclear how to use this coupling constant in time-dependent problems. 
In what follows, we propose a modified form of $g$ that circumvents these limitations.

{\color{black}
{\it Framework: Coupling constant.} }
For $W=0$, the bosonic system forms a droplet whose size, $R^2$, is exponentially smaller than that of the corresponding two-body bound state, with the exponent set by the particle number $N$~\cite{HammerPRL2004}. Consequently, the coupling $g$ in free space is mainly determined by the value of $R^2$. 

The size $R$ is the only length scale of the many-body problem that determines the properties of the droplet, e.g., its density is given by  $N|\psi|^2\sim F(r/R)^2/R^2$, where $r=|\mathbf{x}|$; $F$ denotes a universal function that captures the shape of the droplet. The exponential dependence of the parameter $R$ on $N$ and the fact that the amplitude of the function $\psi$ is given by $1/R$ motivates the following ansatz for $g$ {\color{black}in the limit of large particle numbers, $N\gg 1$,}
\begin{equation}
g\simeq -\frac{4\pi}{\ln\left(\alpha |\psi(\mathbf{x})|^2/ B_2\right)},
\label{eq:g}
\end{equation}
where $\alpha=2.607$. Our choice of the parameter $\alpha$ is fixed by the ground state energy derived in Ref.~\cite{Petrov2024}, as described in the {\color{black}End Matter}. 
For $W=0$, Eq.~\eqref{eq:g} is (within logarithmic accuracy) equivalent to $-4\pi/|\ln(R^2 B_2)|$ derived in Ref.~\cite{HammerPRL2004} everywhere except the points where $\psi(\mathbf{x})$ vanishes. This region, however, does not contribute to the energy functional 
in Eq.~\eqref{eq:energy} where $g$ is multiplied by~$|\psi|^2$.

\begin{figure}[t]
\includegraphics[width=0.48\textwidth]{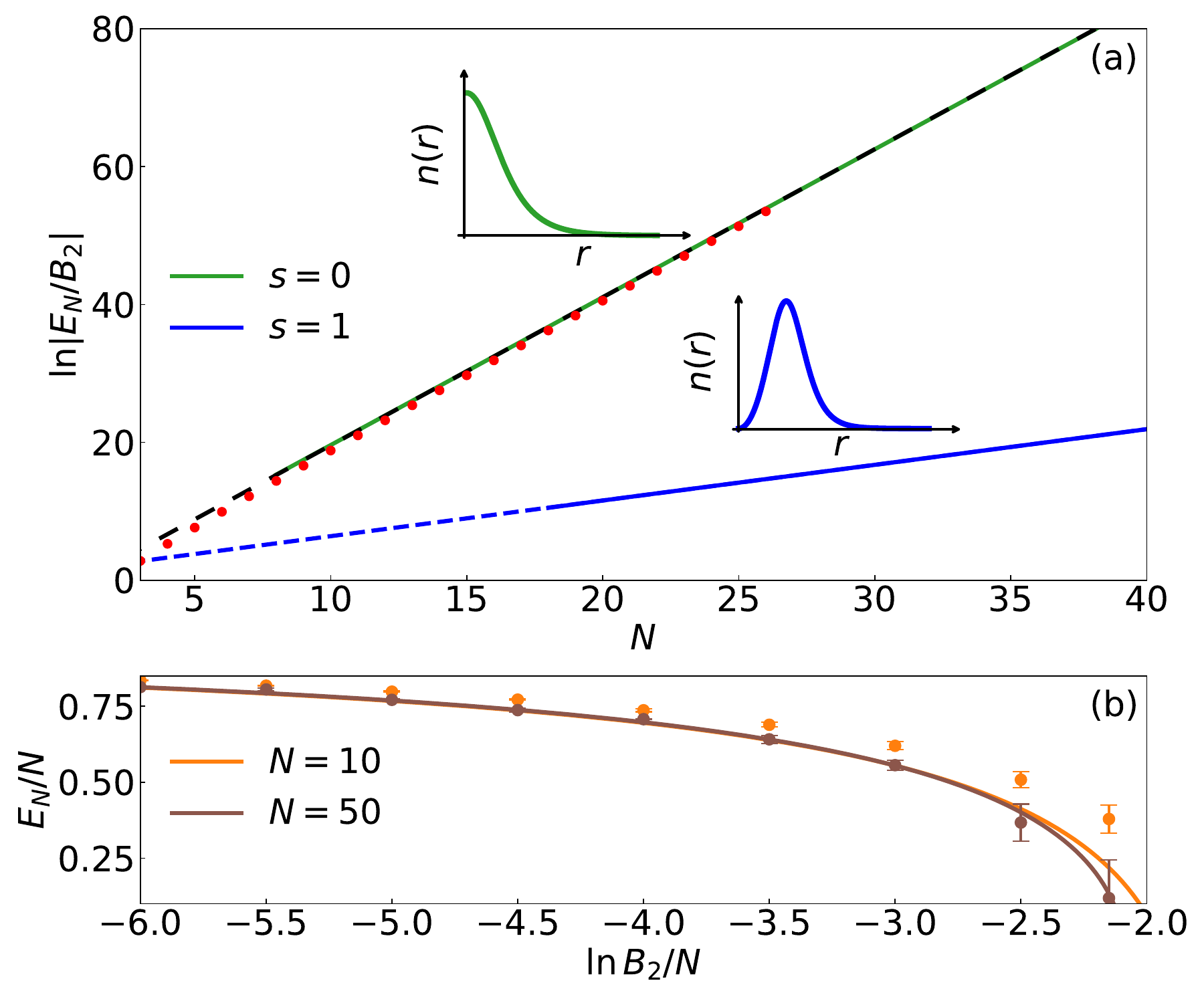}
    \caption{(a) {\color{black}Energy, $\ln \abs{E_N/B_2}$, as a function of the number of bosons $N$ for the lowest energy states without ($s=0$) and with ($s=1$) vorticity for the untrapped system  ($W\to0$)}. The solid green line shows numerical solution of Eq.~\eqref{eq:GPE} for $s=0$; the black dashed line shows the expected asymptote $\ln \abs{E_N/B_2}=\frac{4\pi N}{\mathcal{G}}-1.91$~\cite{Petrov2024}. The red dots demonstrate the numerically exact solution of the many-body problem available for `small' values of $N$~\cite{BazakIOP2018}. The blue line shows the lowest energies for $s=1$. The blue dashed line is a linear fit to this result,  $\ln \abs{E_N/B_2}=0.5174N+1.218$. The insets sketch the radial densities for the states. (b) The lowest energy solution of Eq.~\eqref{eq:GPE} {\color{black}in a trap for weak/intermediate interactions} (solid lines) together with the numerical solution of the many-body problem obtained using flow equations for bosons (dots with error bars), {\color{black}see End Matter for details}.} 
    \label{fig:fig1}
\end{figure}

 Although Eq.~\eqref{eq:g} is inspired by the strongly-interacting regime ($B_2/\hbar\omega\gg1$) where the trap can be neglected, it remains a natural choice for weakly interacting systems in a trap, i.e., when $B_2\to0$. 
 Indeed, in this case, the coupling constant obeys $g\simeq -4\pi/\ln(a^2)$, see, e.g., Ref.~\cite{Tononi2024}. 
 Within logarithmic accuracy, this expression coincides with Eq.~\eqref{eq:g} (recall that $B_2\sim 1/a^2$, see, e.g., Ref.~\cite{Levinsen2015}). 
 Finally, to assess the validity of Eq.~\eqref{eq:g} at intermediate interaction strengths, we perform numerical calculations as presented in Fig.~\ref{fig:fig1}, see the discussion below and Supplemental Material (SM)~\cite{supp}. [Intermediate interaction strengths correspond to $(\ln B_2)/N\simeq- 2.15$~\footnote{$(\ln B_2)/N$ is a parameter that we shall use to quantify the interaction strength. For simplicity, we omit the parentheses in the numerator and write it simply as $\ln B_2/N$ in what follows.}, see  Refs.~\cite{Tononi2024,Brauneis2024} and below.]

{\color{black}
{\it Framework: Gross-Pitaevskii equation.} }{\color{black}Once the coupling constant is established, we use Eq.~\eqref{eq:g} in Eq.~\eqref{eq:energy} and the equation of motion $i  \partial\psi/\partial t = \delta E[\psi]/\delta \psi^*$ to derive the generalized Gross-Pitaevskii equation that governs the physics of the problem}
\begin{equation}
    i  \pdv{\psi}{t} = \qty[ -\frac{1}{2} \pdv[2]{\mathbf{x}}  + W(\mathbf{x})  +  G  N \abs{\psi}^2]\psi,
    \label{eq:GPE}
\end{equation}
where $G=g + g^2/(8 \pi)$~\footnote{It may be tempting to disregard the $g^2$ term due to its typically minor energy contribution, just as it is typically done for repulsive systems~\cite{ChernyPRE2001}. However, in finite systems and time-dependent scenarios, it is essential for preserving energy conservation.}. Equation~\eqref{eq:GPE} provides a powerful framework that greatly facilitates the analysis of 2D bosons with attractive interactions. {\color{black}To illustrate this, we start by calculating static properties of the system}, i.e., we solve Eq.~\eqref{eq:GPE} in polar coordinates using the ansatz $\psi(\mathbf{x},t)=e^{-i\mu t}e^{is\phi}\sqrt{n(r)}$, where $\mu$ is the chemical potential, and the topological quantum number $s$ determines the vorticity of the system. 
{\color{black}This ansatz allows us to recover established results and enables new advances.}

{\color{black}
{\it Benchmark: Strong interactions.}} {\color{black}For sufficiently strong interactions, the trap can be neglected~\cite{Tononi2024,Brauneis2024}, and the system forms a universal many-body bound state.
Given a `weak' (logarithmic)} dependence of $G$ on $\psi(\mathbf{x})$, we gain analytical insight into the equilibrium properties of these states by treating $G$ as constant. Under this assumption, Eq.~\eqref{eq:GPE}  with $W=0$ is stable only for specific values of $G N=\mathcal{G}<0$, where $\mathcal{G}$ is independent of $N$~\cite{bakkalihassani2022}. These values correspond to the fixed points of the energy functional in Eq.~\eqref{eq:energy}, where the {\color{black} the kinetic energy is balanced by particle-particle attraction.} The corresponding solution $\psi$ has the form $\psi\sim e^{is\phi}F_{\mathcal{G},s}(r/R)/R $, here $F_{\mathcal{G},s}$ is a universal function parametrized by $\mathcal{G}$ and $s$. The associated length scale follows from Eq.~\eqref{eq:g} for $N\gg 1$: $R^2\simeq 1/B_2 e^{4\pi N/\mathcal{G}}$.

The smallest possible $|\mathcal{G}|$, $\mathcal{G}\simeq-5.84$, determines the ground state, whose static properties are by now well established~\cite{HammerPRL2004,Petrov2024}. Within our framework, these properties can be readily studied numerically using Eq.~\eqref{eq:GPE}~\footnote{In practice, calculations are performed in the presence of a very weak external potential, which enhances the stability of our numerical approach. See the Supplemental Material for further information about the numerical method~\cite{supp}.}, see Fig.~\ref{fig:fig1}. 
{\color{black}In particular, it is straightforward to recover the universal dependence of energy on the number of particles $E_N\simeq B_2 e^{4\pi N/|\mathcal{G}|}$~\cite{HammerPRL2004}.

Among the most promising experimental platforms for observing this universality are trapped ultracold atomic gases~\cite{Bloch2008}. To guide the corresponding experimental effort, the influence of the trapping potential on the system must be systematically explored. While this investigation is numerically demanding and typically requires advanced computational methods~\cite{Tononi2024}, we demonstrate below that the generalized Gross-Pitaevskii equation enables accurate and computationally efficient analysis of these effects. }

 {\color{black}
{\it Benchmark: Intermediate interactions.} 
To demonstrate the accuracy of the Gross-Pitaevskii equation in accounting for trap effects, we calculate the ground-state energy of the trapped system at intermediate interactions,  i.e., when $|GN|$ is smaller than $|\mathcal{G}|$, which leads to $\ln B_2/N\lesssim 4\pi/\mathcal{G} \simeq -2.15 $.  This ensures the most pronounced interplay between trapping and interparticle interactions, cf. Ref.~\cite{Brauneis2024}. To perform calculations, we employ the flow-equation approach~\cite{Kehrein2006,Tsukiyama2011} (see also End Matter) in the specific form developed for degenerate Bose systems~\cite{Volosniev2017,Brauneis2021,supp}, which not only provide benchmark data but also represent the first application of these equations to two-dimensional systems, thereby opening new avenues for their extension to three dimensions.}

The overall agreement between the data in Fig.~\ref{fig:fig1}(b) is good (see SM~\cite{supp} for the corresponding densities), validating the applicability of the generalized GPE in this regime.
Note that the agreement between the results is better for $N=50$ than for $N=10$ {\color{black} (see SM~\cite{supp} for similar results for other particle numbers)}. This behavior, which mirrors the strongly interacting limit -- see the difference between the red dots and the solid green line in Fig.~\ref{fig:fig1}(a) -- is expected. Our parametrization of the interaction strength, particularly the numerical value of $\alpha$, explicitly relies on the assumption that $N \gg 1$ {\color{black}(see End Matter)}. Nevertheless, the discrepancy is already marginal at $N \sim 10$ and does not affect our main conclusions. {\color{black} Having benchmarked the generalized Gross-Pitaevskii equation against established static results, we now apply it to explore less charted regimes. We begin by investigating trap effects in a regime beyond the reach of the flow equation method.}

\begin{figure}[t]
\includegraphics[width=0.48\textwidth]{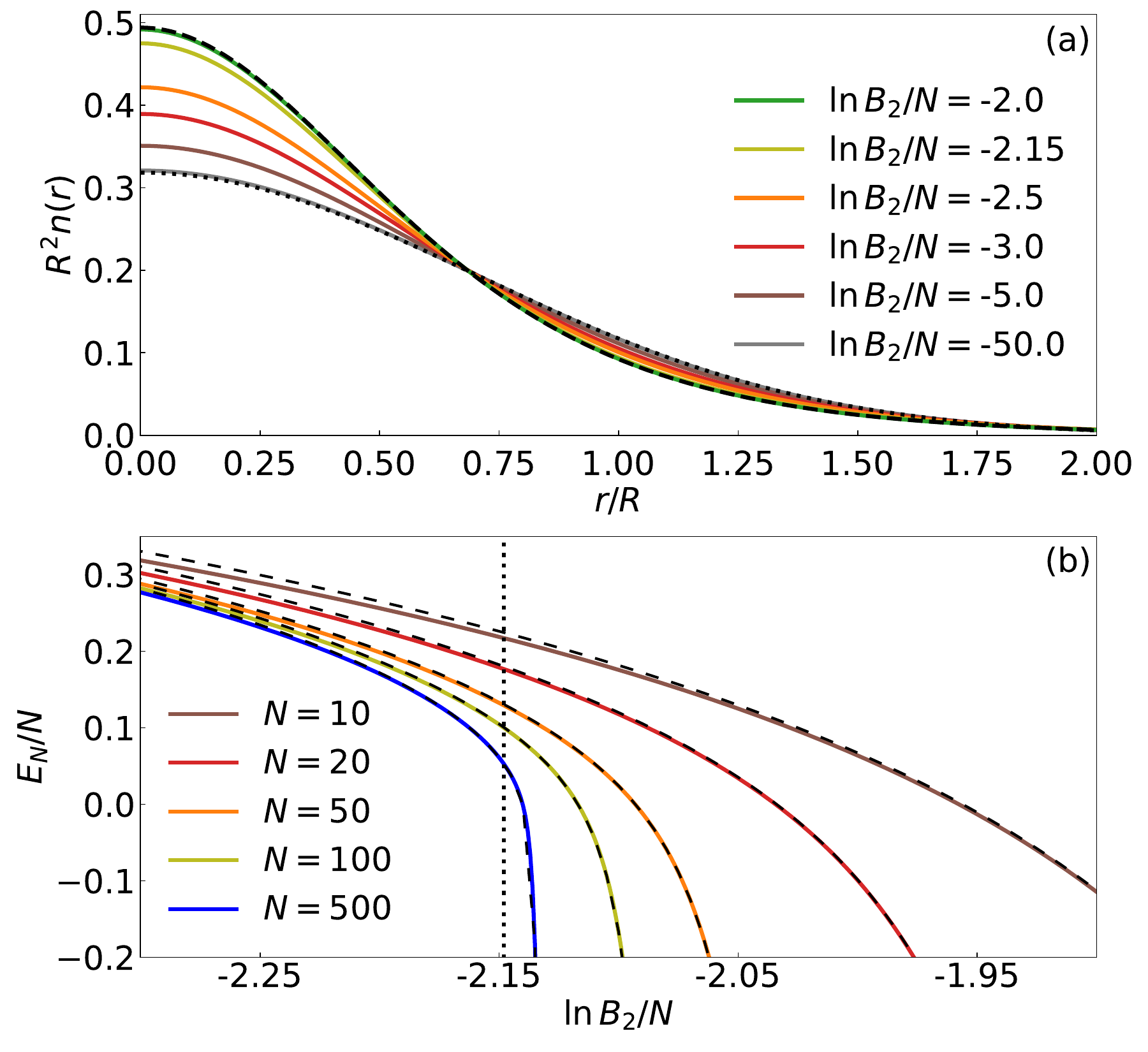} 
\caption{
(a) {\color{black}The density of the trapped Bose gas in the transition between the weakly- and strongly-interacting limits. 
    The dimensionless density, $R^2n(r)$, for the ground state without vorticity ($s=0$) and $N=50$ is plotted as a function of $r/R$ for different interaction strengths $\ln B_2/N$}. The dotted black curve shows the non-interacting harmonic oscillator solution, whereas the dashed curve represents the Townes-soliton profile.
(b) {\color{black}Energy of the Bose gas per particle, $E_N/N$,} as a function of  the interaction parameter $\ln B_2/N$.
{\color{black}The solid curves show the numerical solution to Eq.~(\ref{eq:GPE}), the dashed black curves present the result of minimization of Eq.~(\ref{eq:energy}) with respect to the Townes-soliton profile}. Different curves correspond to different numbers of particles. The vertical dotted line shows the transition point in the limit $N\to\infty$~\cite{Tononi2024,Brauneis2024}.}
    \label{fig:fig3}
\end{figure}

{\it {\color{black} Discussion: Effect of the trap on universal states.}} 
To further investigate the effect of the trap, we solve Eq.~\eqref{eq:GPE} numerically (see SM~\cite{supp} for details). In the weakly interacting regime ($B_2 \to 0$), the density is predominantly shaped by the trapping potential. In contrast, for strong interactions ($B_2 \to \infty$), the universal (so-called Townes soliton~\cite{Chiao1964}) profile is expected to emerge. Figure~\ref{fig:fig3}(a) illustrates the transition between these two limits.  The rescaled density exhibits a crossover behavior, with only a weak dependence on the interaction strength~\footnote{This weak dependence stands in contrast to the behavior observed in a bubble trap~\cite{Tononi2024}, which features a transition from a uniform density to a localized droplet. In that case, the transition involves spontaneous symmetry breaking, potentially leading to dynamics that differ significantly from those presented here. The existence of a vanishing energy Bogoliubov mode in Eq.~(\ref{eq:GPE}) for a bubble trap confirms this assertion. [For a homogeneous density, the Bogoliubov analysis can be performed as in Ref~\cite{Tononi2024}].}.

It is important to note that the generalized Gross-Pitaevskii equation admits solutions for all finite values of $B_2$ and $N$, in contrast to the standard GPE, which exhibits mass concentration (collapse) as the interaction strength approaches a critical value from below~\cite{Guo2013}.
{\color{black}Mass concentration cannot occur in our `asymptotically free' theory, because the coupling constant vanishes at high densities. In other words,  the physics of the high-density regime is governed primarily by kinetic energy, which does not allow the system to collapse.

 Instead of the collapse, in the generalized GPE, the energy shows a smooth increase, which becomes rapid} near the critical parameter value~$\ln B_2/N\simeq -2.15$, as shown in Fig.~\ref{fig:fig3}(b). 
 The derivative of the energy also varies smoothly (see SM~\cite{supp}), suggesting a crossover rather than a sharp phase transition.
{\color{black}Within our framework, the crossover also manifests in other system properties, such as the breathing frequencies, $\Omega$.}

\begin{figure}[t]
\includegraphics[width=0.48\textwidth]{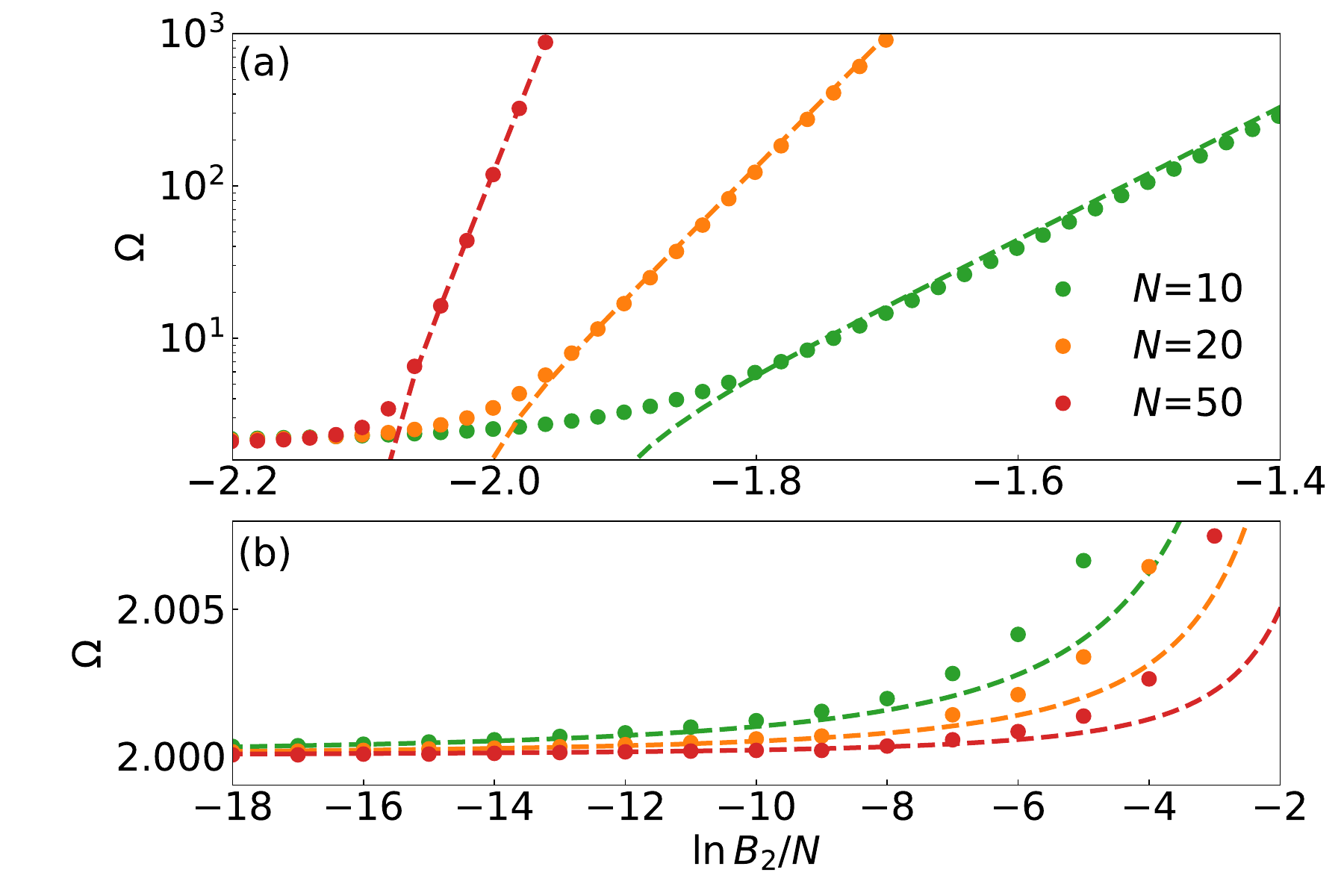} 
    \caption{{\color{black}Breathing frequencies $\Omega$ of a weakly-perturbed trapped Bose gas} as a function of the interaction strength $\ln B_2/N$ for different numbers of bosons $N$ in (a) strong and (b) weak interaction regimes. Dots demonstrate frequencies from the numerical solutions. The dashed curves show analytical results valid for $N\gg1$: (a) $3.8|E_N|/\sqrt{N}$; (b) $2+N/(\ln B_2)^2$.}
    \label{fig:fig2}
\end{figure}

{\color{black} {\it Discussion: Breathing dynamics.} To perform a numerical analysis of the breathing dynamics, we consider the time evolution of the system after a sudden change of the interaction from $\ln B_2/N$ to $\ln B_2/N -0.001$.  The resulting frequency is presented in Fig.~\ref{fig:fig2}.} We see the crossover in the vicinity of $\ln B_2/N\simeq -2.15$ where the frequency of oscillations starts to rapidly increase. This increase becomes sharper as the number of particles increases, reflecting the results presented in Fig.~\ref{fig:fig3}(b).

The generalized Gross-Pitaevskii equation enables an analytical calculation of the frequency {\color{black}(see End Matter)}, thereby providing insight into the system’s limiting behavior.
For weak interactions, we derive $\Omega\simeq 2+N/(\ln B_2)^2$~\footnote{\textcolor{black}{For an experimental validation of this result, the quasi‑2D nature of realistic cold‑atom setups must be taken into account. Effects arising from the quasi‑2D geometry also break the SO(2,1) symmetry, leading to additional corrections to the breathing‑mode frequency~\cite{Olshanii2010,Hu2011}. These effects are, however, extrinsic and can be disentangled from the intrinsic symmetry breaking discussed here by varying the trapping frequency along the tightly confined direction.}}.  
{\color{black}The deviation of the breathing frequency from $\Omega=2$, characteristic of the SO(2,1) symmetry that arises from scale invariance in two dimensions~\cite{Pitaevskii1997}, is a manifestation of `quantum anomaly' captured in our framework via a density-dependent coupling constant.}
For strong interactions,  we derive a trap-independent frequency $\Omega \simeq 3.8 |E_N|/\sqrt{N}$. This expression is in excellent agreement with recent free-space calculations by Petrov~\cite{Petrov2024}, further validating the generalized GPE. 

\begin{figure}[t]
\includegraphics[width=0.48\textwidth]{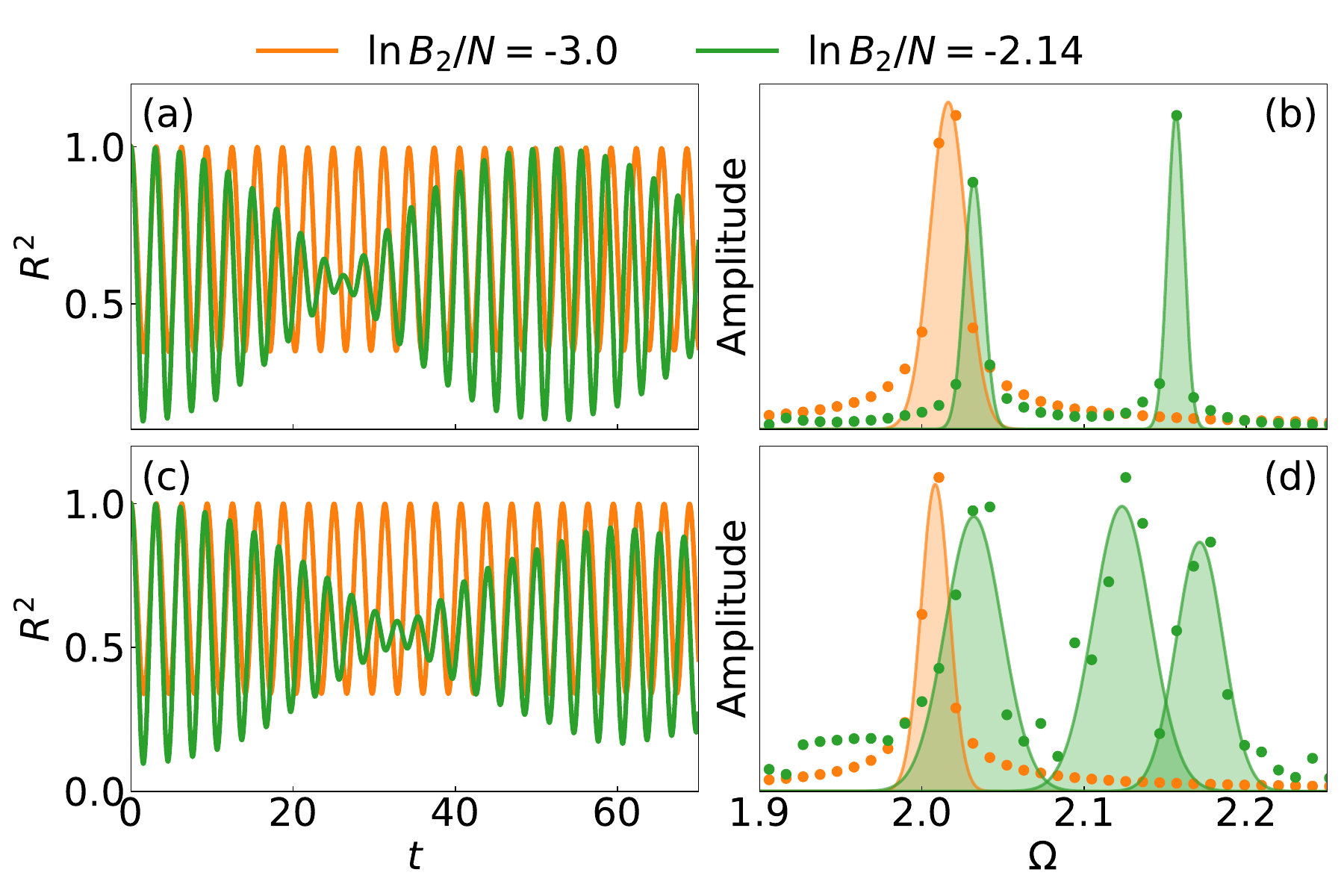} \\
   \caption{{\color{black}Quench dynamics for a trapped Bose gas in the vicinity of the critical interaction strength} with $N=20$ (a-b) and $50$ (c-d) from a non-interacting system to the interaction strength $\ln B_2/N=-3.0$ and $-2.14$. Panels (a,c) demonstrate the size of the system, $R^2$, as a function of time $t$. Panels (b,d) demonstrate the Fourier transforms of the $R^2$ signal (dots). To facilitate interpretation, we fit each peak with a Gaussian function (solid lines with shaded areas).}
   \label{fig:figure4}
\end{figure}

{\it Summary:} In this Letter, we introduced and explored a non-linear equation for attractive 2D Bose systems, 
providing a bridge to the well-established mean-field description of repulsive interactions~\cite{LiebCMP2001,LeeJPB2002,ChernyPRE2001}. The generalized GPE enables a quantitative description of the system using simple (semi-)analytical tools. For instance, in the challenging regime of intermediate interaction strengths, the system's energy can be obtained variationally by employing the Townes-soliton profile  $f_\text{T}(r/\tilde R)$ where $\tilde R$~ is a variational parameter that minimizes the mean-field energy (see Fig.~\ref{fig:fig3}(b) and SM~\cite{supp}).

{\color{black}{\it Outlook: Quench dynamics.}} We anticipate that one of the main applications of the generalized Gross-Pitaevskii equation will be the exploration of non-equilibrium dynamics and {\color{black}
quantum control (see, e.g., Ref.~\cite{Hasan2024} for a recent use of an effective mean-field approach in this context)} -- an area that appears otherwise inaccessible for 2D attractive Bose gases. To illustrate this possibility, we consider the quench dynamics of the system, a standard paradigm for investigating non-equilibrium physics in closed systems~\cite{Polkovnikov2011}.
We initiate the system in the ground state of the harmonic oscillator at $t=0$ and consider the time dynamics of $\langle r^2\rangle$ at $t>0$, assuming a finite value of $B_2$, see Fig.~\ref{fig:figure4}.
We see that for weak interactions, the quench dynamics generates oscillations at frequencies $\Omega\simeq 2$, consistent with the scale-invariant dynamics. At stronger interactions, this universality breaks down, and we observe an interference pattern (beating) in the time evolution, {\color{black}see End Matter}.

{\color{black} \textit{Outlook: Universal excited states.}}
Our work motivates further mathematical and numerical studies to validate the proposed framework and identify experimental conditions for observing 2D universality in attractive systems.
A concrete direction involves studying  universal excited states predicted by the generalized GPE~\footnote{Notably, it is already known that the 2D three-body (four-body) system supports a single excited state with zero angular momentum and energy $1.27 B_2$~\cite{Bruch1979,Nielsen1999} ($25.1 B_2$~\cite{Platter2004}). 
According to Eq.~\eqref{eq:GPE} one could expect that the large-$N$ limit for such excited states reads $B_2 e^{4\pi N/|\mathcal{G}|}$, where $|\mathcal{G}| = 38.6$ corresponds to the excited state of the generalized GPE with $s=0$. It is also plausible that these excited states are beyond the reach of our mean-field-like approach. Classifying such states in 2D attractive systems presents a compelling direction for future research.}, which are connected to the spectrum of scale-invariant solutions to the standard 2D GPE, either without~\cite{Haus1966} or with vorticity~\cite{Kruglov1992}, see Ref.~\cite{bakkalihassani2022} for a review.

To illustrate one such state, we considered the vortex solution. Specifically, Fig.~\ref{fig:fig1} presents the energy of the lowest-energy configuration with topological charge $s = 1$. This solution corresponds to a critical interaction strength of $\mathcal{G} \simeq -24.15$. The energies of the corresponding states are $E^e_{N}\sim 1/R^2$; they lead to the universal scaling relation $E^e_{N+1}/E^e_N \simeq 1.7$, which may be contrasted with the ground-state scaling $E^g_{N+1}/E^g_N \simeq 8.6$~\cite{HammerPRL2004}. Note the markedly slower increase in energy with growing particle number -- a distinctive trait of the excited-state spectrum. This suggests that topologically protected excited states may be more suitable for the experimental observation of 2D universality than the ground state. \textcolor{black}{It is important to note that the stability of such vortex states (both in free space and in trapped systems) is an open question with direct relevance to an experiment. We are currently addressing this issue through both the Bogoliubov-de Gennes approach and dynamical perturbation tests. }

\begin{acknowledgments}
    We thank Robert Seiringer for useful discussions and making us aware of Refs.~\cite{rajeev1999a,rajeev1999b}. We thank Dmitry Petrov for enlightening correspondence. We gratefully acknowledge the National Science Centre, Poland (grants no.~2020/38/E/ST2/00564 and 2023/49/N/ST2/03820) for financial support and Poland's high-performance computing infrastructure, PLGrid (HPC Centers: ACK Cyfronet AGH), for providing computer facilities and support (computational grant no.~PLG/2024/017527); A.~G.~V. acknowledges the support from Danish National Research Foundation through the Center of Excellence
``CCQ'' (DNRF152); H.-W.~H. acknowledges the support by Deutsche
Forschungsgemeinschaft (DFG, German Research Foundation) under Project ID 279384907 – SFB 1245,  and by the BMFTR contract No. 05P24RDB.
\end{acknowledgments}

\clearpage

{ \color{black} \section{End Matter} }

\section{Parameter $\alpha$}

\begin{figure}[t]
\includegraphics[width=0.48\textwidth]{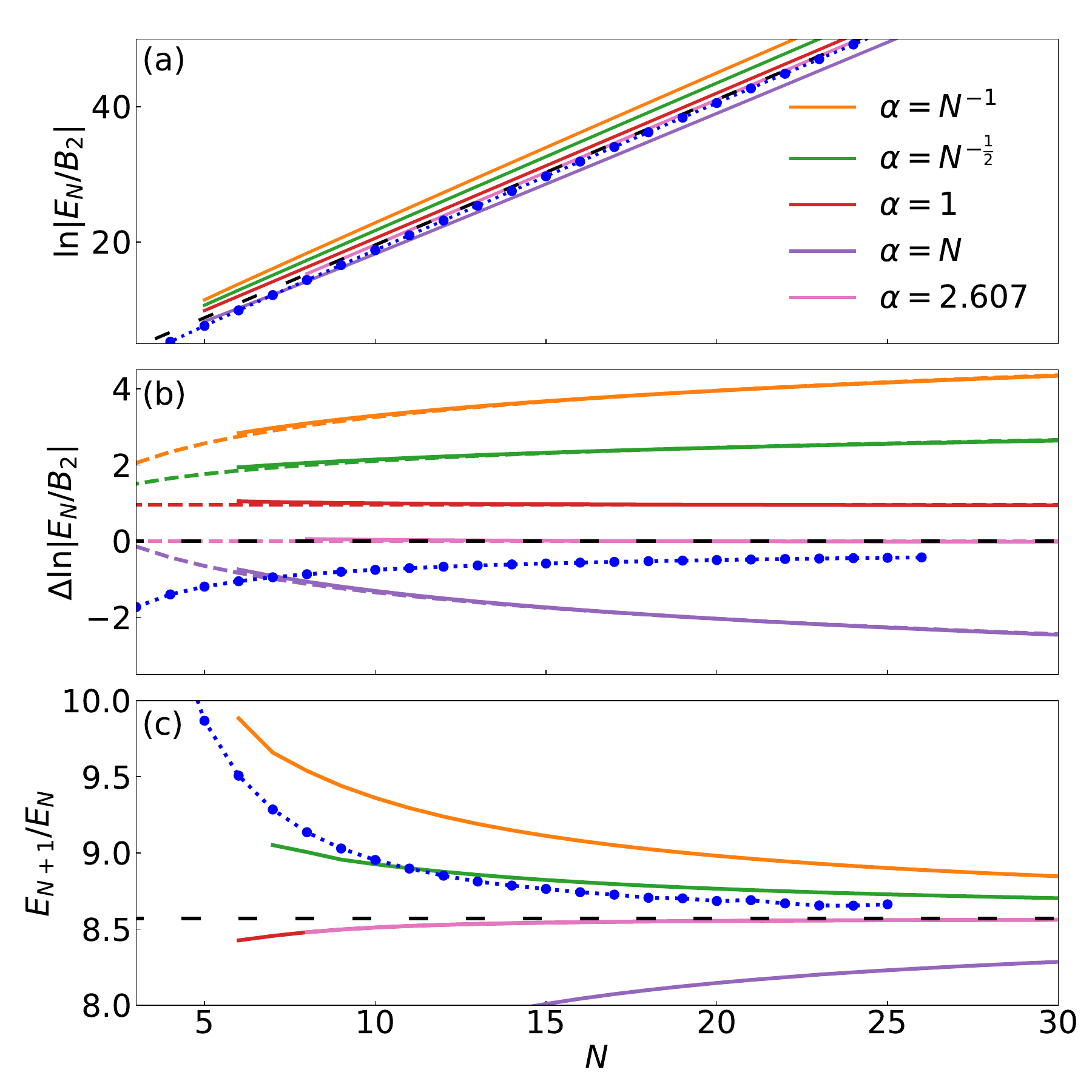} \\
   \caption{ (a) Ground-state energy of the untrapped Bose gas  $\ln |E_N/B_2|$. (b) Energy differences between $\ln |E_N/B_2|$ and the  expected asymptote $\Delta \ln |E_N/B_2|=\ln |E_N/B_2|-(4N/C+c_1)$~\cite{Petrov2024} for the data presented in panel (a). (c) Energy ratios $E_{N+1}/E_N$ for the data presented in panel (a). Solid lines represent solutions of the generalized GPE. The black dashed line represents the solution of Ref.~\cite{Petrov2024}, and the blue dots with a dotted line show the results of Ref.~\cite{BazakIOP2018}. In panel (b), the dashed colored lines show $\Delta \ln |E_N/B_2|$ where $E_N/B_2$ is calculated from Eq.~\eqref{eq:BN_alpha} with $\kappa=1.448$. }
   \label{fig:sm_figure1}
\end{figure}

To determine $\alpha$ in Eq.~(\ref{eq:g}) from the Main Text, we first compute the energies of the system in the limit $W\to 0$, where according to Ref.~\cite{Petrov2024} the ground state energy for $N\gg 1$ approaches $B_N^{s=0} \to B_2 e^{\frac{4N}{C}+c_1}$ with $C\simeq 1.862$ and $c_1\simeq -1.91$. By trial and error, we found that $\alpha\simeq 2.607$ reproduces this limiting behavior well, see Fig.~\ref{fig:sm_figure1}(b). Other values of $\alpha$ can be used if one finds this necessary.
In particular, our model does not rule out the possibility for $N$-dependent $\alpha$: $\alpha(N)= \alpha_0+\alpha_1N^\gamma+... \;$. The inclusion of this $N$-dependence might be needed to describe the energy in the few-body limit accurately. Below, we outline the basic intuition behind determining the value of $\alpha$ that can be used for a further exploration of the generalized GPE.

In the limit $W \to 0$, $N\to\infty$, the many-body solution is well represented by the Townes soliton  $\psi_\text{T}(r)=(2 \pi C \mathcal{R}^2)^{-1/2} f_\text{T}(r/\mathcal{R})$, where $\mathcal{R}^2=\frac{C}{M_2}\ev{r^2}{\psi_\text{T}}$ with $M_2 \simeq 2.211$~\cite{petrov_r}. For an illustration of  $f_\text{T}$, see the Supplemental Material~\cite{supp}. Therefore, the interaction strength in this limit should have the form
\begin{equation}
    g= -\frac{4 \pi}{\ln (\frac{\alpha(N)}{2 \pi C} \abs{f_\text{T}}^2 \frac{1}{\mathcal{R}^2 B_2})}.
\end{equation}
We rewrite this expression at the fixed point of the generalized GPE (note that $G\simeq g$ in the considered limit):
\begin{equation}
    \mathcal{G}\simeq-\frac{4 \pi N }{\ln (\frac{\alpha(N)}{2 \pi C} F_\text{eff} \frac{1}{\mathcal{R}^2 B_2})},
    \label{eq:geff}
\end{equation}
where $\mathcal{G} = -\pi C$ and $F_\text{eff}$ is a numerical constant representing some effective value of the density.  System's energy is $E_N=-\frac{C}{8 \mathcal{R}^2}$ (see, e.g., Ref.~\cite{Petrov2024}) and therefore
\begin{equation}
    \frac{E_N}{ B_2} = -\frac{\pi C^2}{4} \frac{1}{\alpha(N)} \frac{1}{F_\text{eff}} e^\frac{4 N}{ C }.
    \label{eq:BN_alpha}
\end{equation}
Now, we propose that $F_\text{eff}=\kappa \abs{f_\text{T}(0)}^2$ with scaling coefficient $\kappa$. 

To determine $\kappa$ and illustrate the effect of possible dependence of $\alpha$ on $N$, we solve  Eq.~(\ref{eq:GPE}) numerically, for $\alpha(N)=N^{\gamma}$ with $\gamma=-1,0,\frac{1}{2},1$. We use calculated values of $E_N/B_2$ and Eq.~\eqref{eq:BN_alpha} to fit the parameter $\kappa=1.448$. Note that the parameter $\kappa$ is independent of the $\alpha(N)$. In panel (a) and (c) of Fig.~\ref{fig:sm_figure1}, we show $\ln |E_N/B_2|$ and $E_{N+1}/E_N$ for different $\alpha(N)$ compared to Refs.~\cite{Petrov2024, BazakIOP2018}. Panel (b) presents the difference $\Delta \ln |E_N/B_2|=\ln |E_N/B_2|-\qty(4N/C+c_1)$ where $E_N/B_2$ is calculated from the generalized GPE using Eq.~\eqref{eq:BN_alpha} with $\kappa=1.448$.

\begin{figure}[t]
\includegraphics[width=0.48\textwidth]{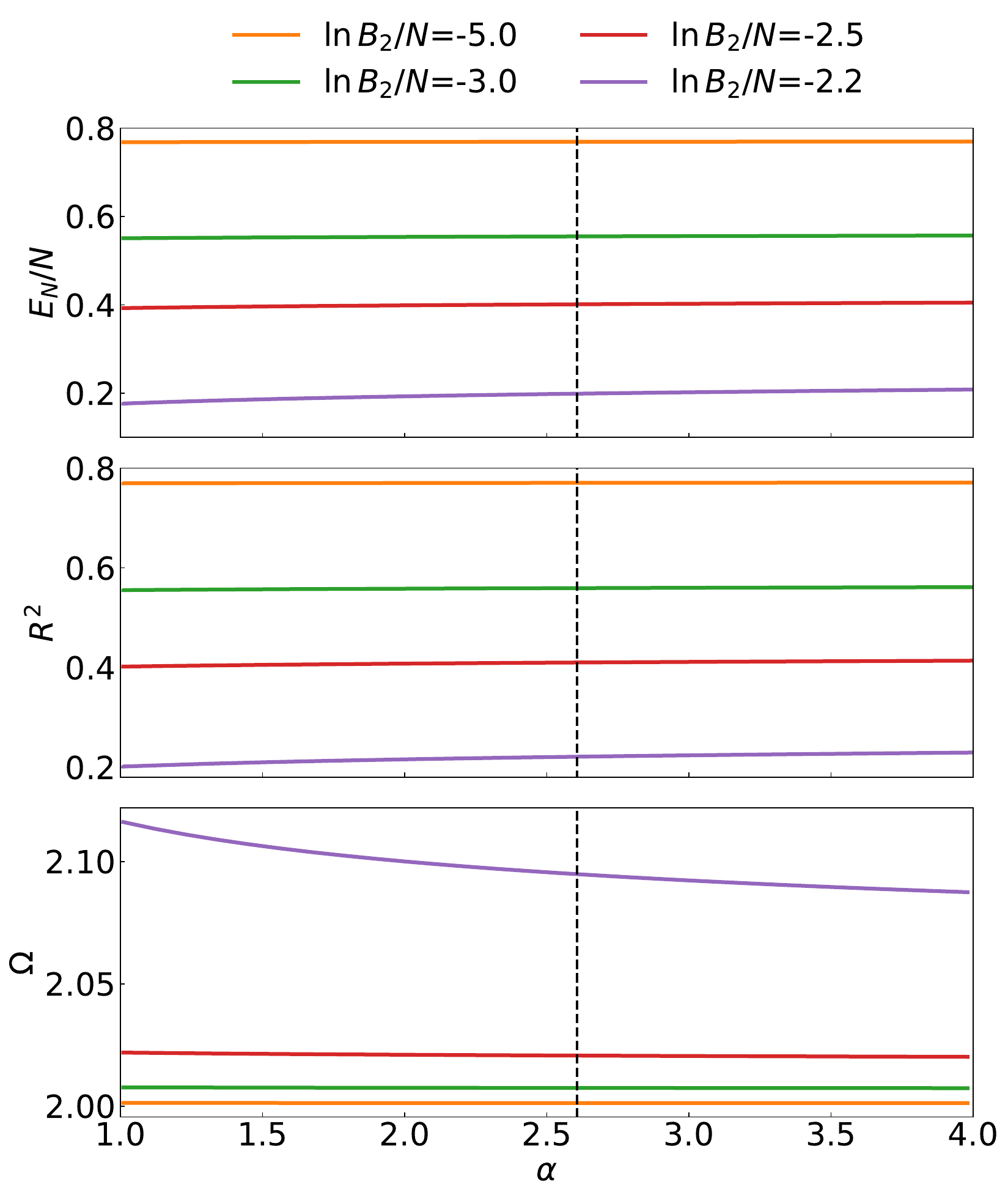} \\
   \caption{ \textcolor{black}{(a) The ground-state energy $E_N$ (b) the size of the system $R^2$ and (c) the breathing mode frequency $\Omega$ as a function of the parameter $\alpha$ for different interaction strength $\ln B_2/N$ for $N=50$. Black vertical dashed line shows value of $\alpha=2.6073$.} }
   \label{fig:new_figure_b}
\end{figure}

Once $\kappa$ is determined, we can estimate the parameter $\alpha(N)$ as
\begin{equation}
    \alpha(N) = - \frac{\pi C^2 }{4} \frac{1}{\kappa \abs{f_\text{T}(0)}^2} \frac{B_2}{E_N} 
 e^\frac{4N}{C}.
\end{equation}
According to Ref.~\cite{Petrov2024}, $E_N = -B_2 e^{\frac{4N}{C}+c_1}$. Then
\begin{equation}
    \alpha(N) = \alpha = \frac{\pi C^2}{4\kappa \abs{f(0)}^2} e^{-c_1} \simeq 2.6073 ,
\end{equation}
where we have used $f_\text{T}(0)=2.207$, see, e.g., Ref.~\cite{HammerPRL2004}. Results for $\alpha=2.6073$ are also presented in Fig.~\ref{fig:sm_figure1}. \textcolor{black}{In Figure \ref{fig:new_figure_b}, we show how the small variation of the parameter $\alpha$ affects energy $E_N$, system's size $R^2$, and breathing mode frequency $\Omega$. In the weakly-interacting regime, the properties are almost independent of $\alpha$. The impact becomes more significant near the transition to the strongly-interacting regime, where small variations of $\alpha$ can strongly affect the results.} 

In a more general approach, 
\begin{equation}
    E_N = -B_2 e^{\frac{4N}{C} + \theta(N)},
    \label{eq:sup_en}
\end{equation} 
where $\theta(N)=\sum_{i=1} c_i N^{\gamma_i}$. Parameters $c_i$ and $\gamma_i$ may be fitted to the results of other methods, see Ref.~\cite{BazakIOP2018} for an example of $\theta(N)$ fitted to the few-body limit results. Equation~\eqref{eq:sup_en} leads to the expression
\begin{equation}
    \alpha(N) = \frac{\pi C^2}{4\kappa \abs{f_\text{T}(0)}^2}  e^{-\theta(N)} ,
  \end{equation}  
where properly fitted $\theta(N)$ may improve the accuracy of generalized GPE in the few-body limit.

\section{Flow equations}

In the flow-equation method, we block-diagonalize the Hamiltonian $H$ using the flow equation
\begin{equation}
    \dv{H}{s} = \qty[ \eta, H],
\end{equation}
where $s$ is the flow parameter (which can also be understood as an imaginary time). With the proper choice of the generator of the flow $\eta$, the off-diagonal matrix elements will vanish in the $s \to \infty$ limit.
The Hamiltonian of the system reads  
\begin{equation}
H=\sum_{i=1}^N\left(-\frac{\hbar^2}{2m} \pdv[2]{\mathbf{x}_i} +W(\mathbf{x}_i)\right) + \frac{1}{2}\sum_{i,j=1}^N V(\mathbf{x}_i-\mathbf{x}_j),
\end{equation}
where $V$ describes short-range particle-particle interactions; $\{\mathbf{x}_i\}$ are the coordinates of the bosons. We provide additional details in the Supplemental Material~\cite{supp}.

\section{Breathing dynamics}

To calculate breathing-mode frequencies, we first compute the time dynamics of $R^2=\langle r^2\rangle$ (see Supplemental Material~\cite{supp} for detailed derivation). The corresponding equation for large values of $N$ is given by
\begin{equation}
\frac{\mathrm{d}^2 R^2 }{\mathrm{d} t^2}=4 \frac{E_N}{N} - 4  R^2 + 2 N  \ev{ n^2 \frac{\mathrm{d}g}{\mathrm{d}n} }.
\label{eq:r2}
\end{equation} 
This equation is general in the sense that it applies to any time dynamics of the state. For example, a time-independent problem requires
$E_N=N R^2-\frac{N^2}{2} \ev{  n^2 \frac{\mathrm{d}g}{\mathrm{d}n} }$. 
Note that the total energy comprises two contributions: one from the harmonic confinement and another from density-dependent interactions. These contributions are balanced when $E_N=0$, offering an intuitive explanation for the transition point between weakly and strongly interacting regimes, as illustrated in Fig.~\ref{fig:fig3}(b).

If $g$ were a constant, the last term in Eq.~\eqref{eq:r2} would disappear, resulting in interaction-independent dynamics governed solely by the first two terms~\cite{Pitaevskii1996,Pitaevskii1997}. A density-dependent $g$ breaks this picture, leading to a `quantum anomaly',  similar to dimensional transmutation in QCD \cite{Gross:1973id,Politzer:1973fx} or the Efimov effect \cite{Efimov:1970zz,Braaten:2004rn}. {\color{black}This symmetry breaking is an intrinsic property of two-dimensional systems with short-range interactions, contrasting with the extrinsic symmetry breaking induced by a `compactified' third dimension in cold-atom experiments~\cite{Hu2011}.}

We investigate `quantum anomaly' in the breathing dynamics, assuming that the solution has the form $\psi(r/R(t))/R(t)$. The time dynamics of $R(t)$ follows from Eq.~\eqref{eq:r2}:
$\ddot{R}^2 (t)=4E_N/N - 4  R^2(t) + N\overline{\langle n g^2/2\pi\rangle}/R^2(t)$, where the overline highlights that the expectation value is taken with respect to a re-scaled function $\psi(r)$.
Looking for small oscillations of $R(t)$ in the vicinity of the equilibrium value $R(t=0)$, we find the frequency of the breathing mode ($N\gg1$)
\begin{equation}
\Omega=\sqrt{8-4E_N/(NR^2(t=0))},
\label{eq:omega_main}
\end{equation}
where we have used the time-independent solution to Eq.~\eqref{eq:r2}. \textcolor{black}{This expression agrees with the sum-rule approach result, see SM~\cite{supp}.}

Equation~(\ref{eq:r2}) can also be used to analyze quench dynamics presented in Fig.~\ref{fig:figure4}. In particular, we can conclude that the beating in panel (c) arises from the time dynamics of $g$ in Eq.~\eqref{eq:r2}, which resonates with the universal, scale-invariant dynamics. 

\bibliographystyle{apsrev4-2}
\bibliography{apssamp}

\clearpage 

\onecolumngrid
\begin{center}
\textbf{\large Supplemental Material}
\end{center}

\renewcommand{\thefigure}{S\arabic{figure}}
\setcounter{figure}{0}

\renewcommand{\theequation}{S\arabic{equation}}
\setcounter{equation}{0}

\renewcommand{\thesection}{S\arabic{section}}
\setcounter{section}{0}

\makeatletter
\setcounter{secnumdepth}{3}
\makeatother

\twocolumngrid

\section{The flow equation approach (IM-SRG)}

\begin{figure*}[t]
\includegraphics[width=0.96\textwidth]{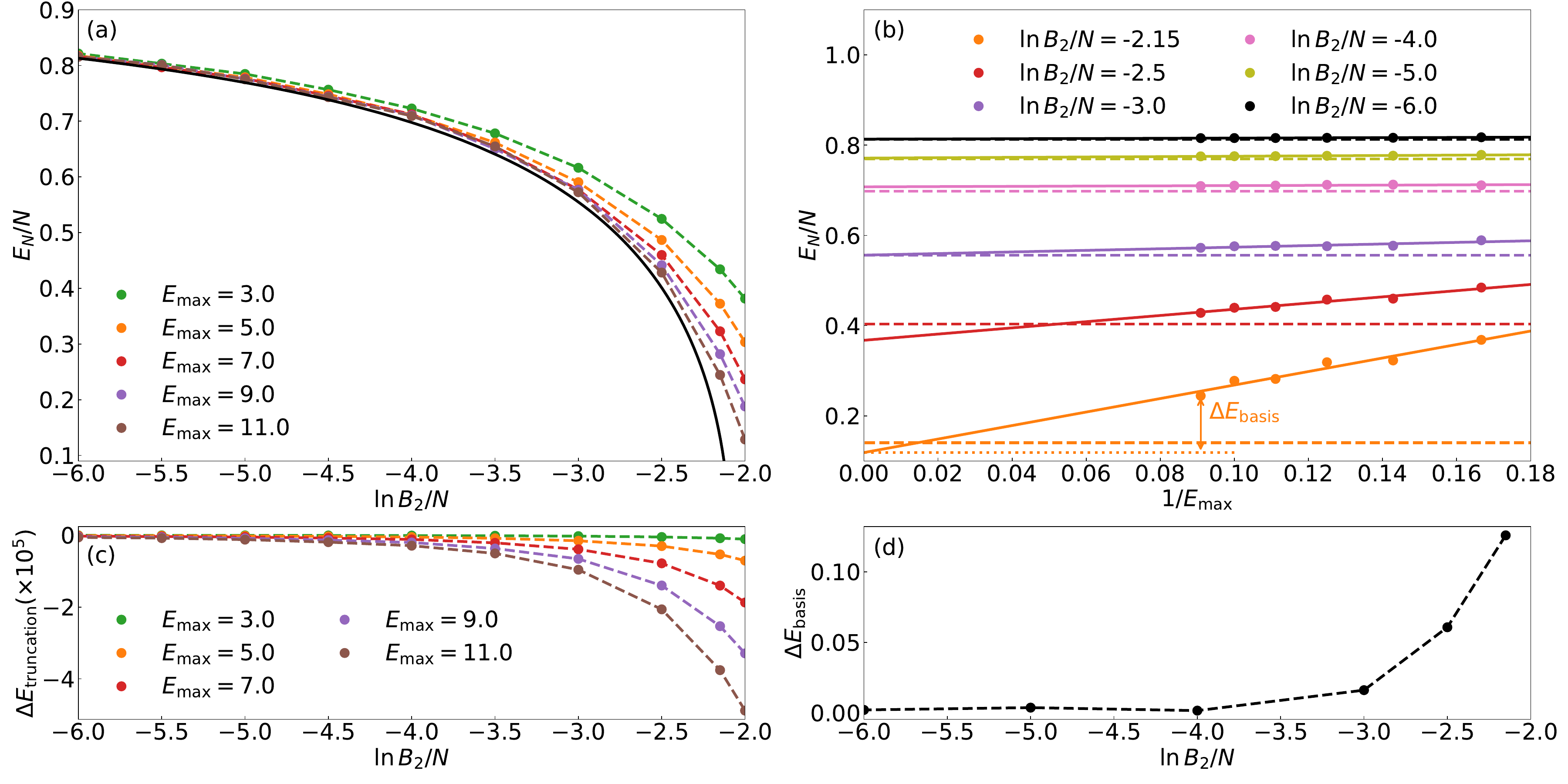} \\
   \caption{The flow-equation approach for $N=50$. (a) Energy $E_N$ as a function of interaction strength $\ln B_2 /N$ for different basis sets parametrized by the maximal allowed energy, $E_\text{max}$, (dots with dashed lines). The solid black curve shows the energy $E_N$ calculated using Eq.~(3) of the main text. (b) Energy $E_N$ as a function of $1/E_\text{max}$ for different $\ln B_2/N$ (dots). Solid lines show linear fit $a_1+\frac{a_2}{E_\text{max}}$. Dashed lines show energies $E_N$ from Eq.~(4). For $\ln B_2/N=-2.15$, we show how the error $\Delta E_\text{basis}$ is calculated. (c) Truncation error $\Delta E_\text{truncation} $ as a function of the interaction strength $\ln B_2/N$ for different $E_\text{max}$. (d) Finite basis set error $\Delta E_\text{basis}$ as a function of the interaction strength.}
   \label{fig:sm_figure2}
\end{figure*}

\begin{figure}[t]
\includegraphics[width=0.48\textwidth]{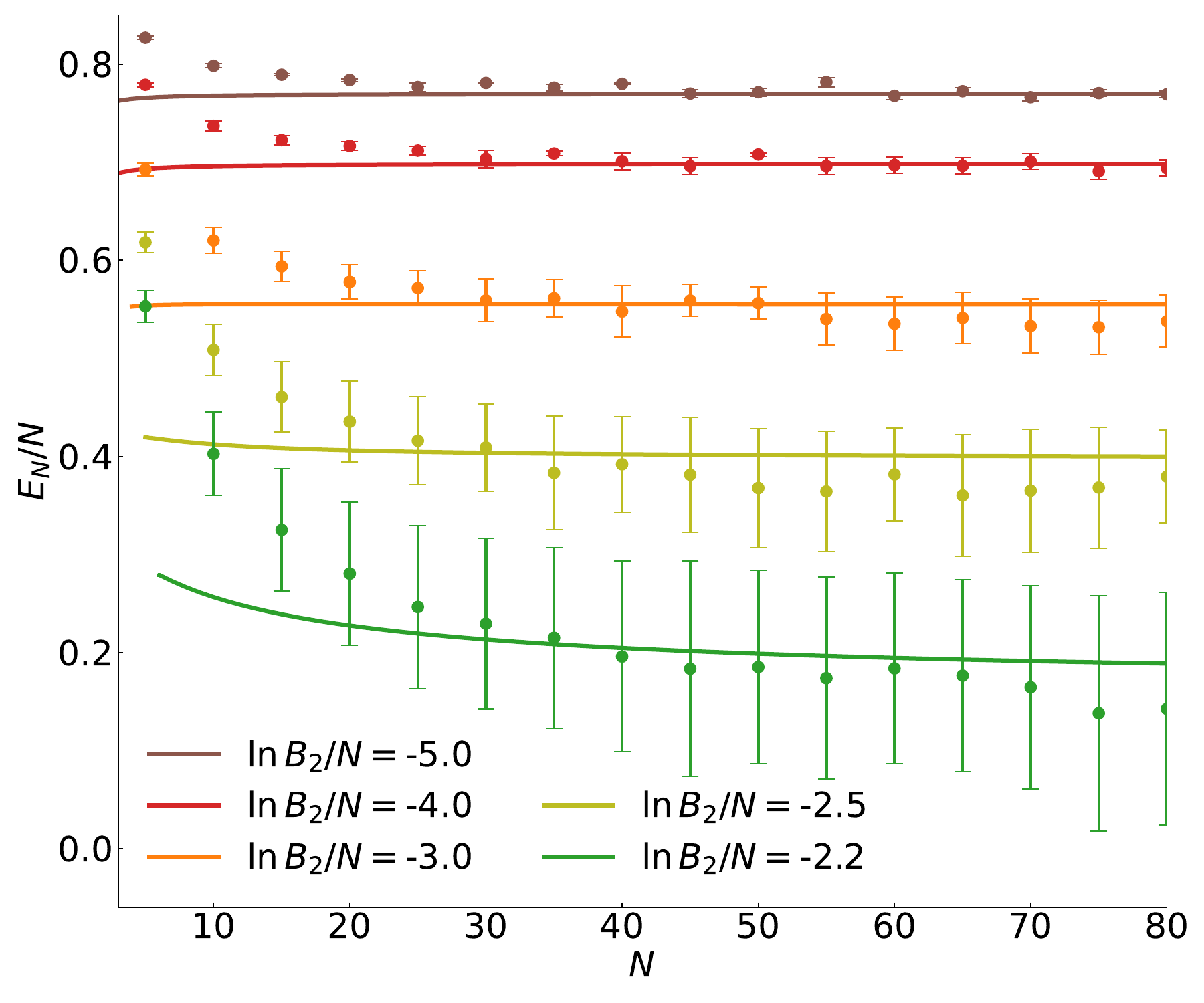} \\
   \caption{\textcolor{black}{The ground-state energy $E_N$ as a function of number of bosons $N$ for different interaction strength $\ln B_2/N$. The solid lines show the energy $E_N$ calculated using Eq.~(3) of the main text. Dots represents results of the flow equations approach with error $\Delta E_\text{basis}$ coming from finite basis size.}}
   \label{fig:sm_figure_a}
\end{figure}

\begin{figure}[t]
\includegraphics[width=0.48\textwidth]{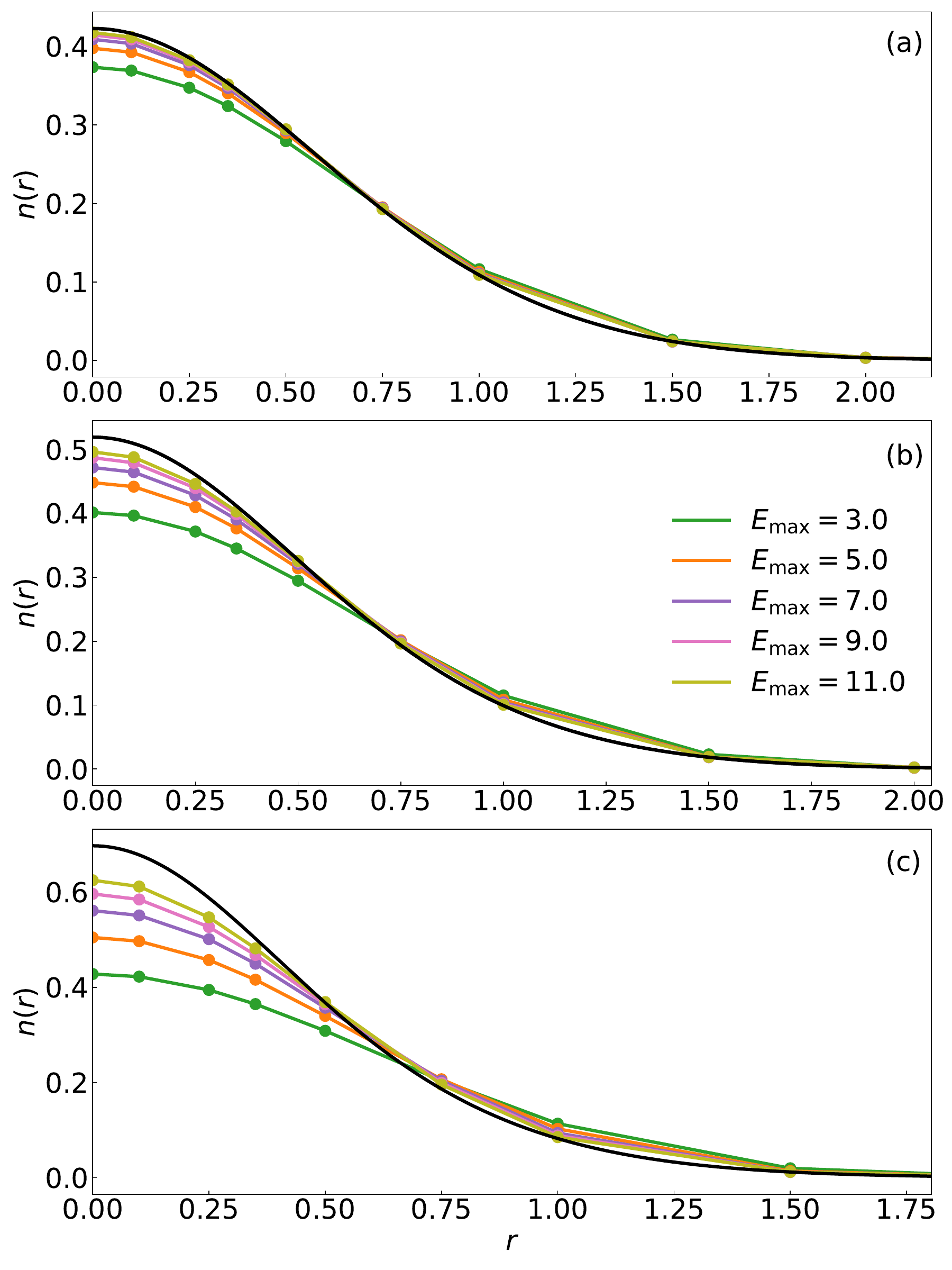} \\
   \caption{Densities $n(r)$ as a function of $r$ for $N=50$ and $\ln B_2/N=$ (a) $-6.0$, (b) $-4.0$ and (c) $-3.0$. The dots with solid lines show results from the IM-SRG method with different $E_\text{max}$. Solid black curves show densities calculated from Eq.~(3). }
   \label{fig:sm_figure3}
\end{figure}

In this section, we discuss the general idea behind the flow-equation approach~\cite{Kehrein2006}, also called the IM-SRG method~\cite{Tsukiyama2011}. For more details and previous applications to one-dimensional bosonic systems, see Refs.~\cite{Volosniev2017,VolosnievPRA2017,Brauneis2021,BrauneisNJP2022,Brauneis2023}.

For our system, the Hamiltonian in canonical quantization using the basis of non-interacting harmonic oscillator states $\{\psi_{(n_r,m)  } \}$ reads as follows 
\begin{equation}
    H = \sum_i \epsilon_i a_i^\dagger a_i + \sum_{ijkl} B_{ijkl} a_i^\dagger a_j^\dagger a_k a_l,
\end{equation}
where the lower index denotes quantum numbers $(n_r,m)$ with one-body energy $\epsilon_{(n_r,m)}=\hbar \omega(2n_r+\abs{m}+1)$ and 
\begin{equation}
    B_{ijkl} = \widetilde{g} \int \dd \mathbf{x} \psi_i(\mathbf{x}) \psi_j(\mathbf{x}) \psi_k(\mathbf{x}) \psi_l(\mathbf{x}),
\end{equation}
where $\widetilde{g}$ is the coupling constant.

The goal of the method is to decouple the ground state of the system from the rest of the Hilbert space. To achieve this, the Hamiltonian is transformed so that the system is expressed in terms of one- and two-body excitations relative to a chosen reference. In our case, the reference state is a non-interacting ground state, corresponding to the lowest-energy basis state $\psi_{(0,0)}$. Alternatively, one may employ a mean-field solution or construct it iteratively, updating the reference state after each flow step. 

During the flow evolution, only one- and two-body operators are kept. The corresponding error $\Delta E_\text{truncation}$ coming from discarding higher-order operators can be estimated using perturbation theory~\cite{Volosniev2017}. Moreover, the second type of error comes from a finite basis set. We limit the basis set to states $\psi_{(n_r,m)}$ with one-body energies $\epsilon_{(n_r,m)}<E_\text{max}$. We fit the energies $E_N$ for each interaction strength $\ln B_2/N$ to the function $a_1+\frac{a_2}{E_\text{max}}$ so that the energy $E_N \eval_{E_\text{max} \to \infty}  = a_1$. We estimate the error due to finite basis set as $\Delta E_\text{basis} = \frac{1}{N} \qty [ E_N(E_\text{max}=11)-a_1 ]$. 

In Figure~\ref{fig:sm_figure2}, in panel (a), we show energies $E_N$ for different sizes of the basis set as a function of $\ln B_2 /N$. Panel (b) highlights the fitting procedure, showing $E_N$ as a function of $E_\text{max}$ together with the linear fit $a_1+\frac{a_2}{E_\text{max}}$. Panels (c) and (d) show errors coming from truncations during the flow evolution and from the finite basis set. Note that the truncation error is a few orders of magnitude smaller than the one coming from the finite basis set. For this reason, in Fig.~1 in the main text, we show only the value of $E_N =a_1$ with error bars given by $\Delta E_\text{basis}$. \textcolor{black}{ Similarly, in Figure~\ref{fig:sm_figure_a}, we show energy $E_N=a_1$ with error bar associated with finite basis size as a function of number of bosons in the system $N$ for different interaction strengths. In a weakly interacting regime for $N>20$, we observe that the IM-SRG method agrees well with the generalized GPE. For strong interactions, truncation of the basis size introduces large error bars for IM-SRG numerical data, which complicates the comparison of the methods.} 
Finally, Fig.~\ref{fig:sm_figure3} shows the ground-state density for different $E_\text{max}$ compared to the densities calculated using Eq.~(3) of the main text.

To compare IM-SRG results to GPE solutions, it is crucial to connect the interaction strength $\widetilde{g}$ to the two-body binding energy $B_2$. To do so, we calculate the lowest energy, $E_2$, of the two-body problem for different values of $\widetilde{g}$ in a finite basis size. This provides us with the function $\widetilde{g}(E_{\mathrm{max}},E_2)$. In IM-SRG calculations, we use 
$\widetilde{g}(E_{\mathrm{max}},B_2)$, which fixes the lowest two-body binding energy to the correct value, see Ref.~\cite{Brauneis2025} for more details.

\section{Numerical method for GPE}

In this section, we discuss the numerical method employed to solve Eq.~(3) of the main text in the stationary and time-dependent versions.

\textit{Boundary conditions.} We utilize the rotational symmetry of the problem and solve the equation for a radial function $f(r)=\sqrt{n(r)}$  on the spatial grid with size $N_\text{g}+1$ with spacing $\dd r$, where we note $f_i \equiv f( r_i )$, and $r_i=i \dd r$. For $s=1$, we expect that $f(0)=0$, which leads to the simple boundary condition $f_0=0$ with artificial points $f_{-1}=0$ and $f_{-2}=0$, ensuring proper treatment of the derivative terms within the 4th order finite difference approximation. For $s=0$, we introduce boundary conditions $\pdv{f}{r} \eval_{r=0} =0$ and $f(0) \neq 0$. To approximate the kinetic operator, we assume a  symmetrical solution using artificial points $f_{-1}=f_1$, $f_{-2}=f_2$, and expand $f(r)$ around $r=0$
\begin{equation}
    f(r) \eval_{r \to 0} \approx f(0) + \frac{r^2}{2} \pdv[2]{f(r)}{r} \eval_{r=0},
\end{equation}
to properly address the kinetic operator at $r_0$ and $r_1$. Additionally, for all $s$, we use boundary condition at $r \to \infty $ as $f \to 0$ by assuming that $f_{N_g+1}=0$ and $f_{N_g+2}=0$.

\textcolor{black}{At this point, we would like to clarify the GPE behaviour in the regions where the solution behaves as $f(r)\to0$, e.g., at $r\to\infty$ and at the vortex core for $s>0$. The full form of the interaction coupling has a well-defined behavior $\frac{n}{\ln n} \to 0$  in this limit, which ensures that no additional renormalisation or numerical cutoff procedure is needed to keep the stability of the solution. }

\textit{Ground-state calculations.} To find the ground state of the system, we employ imaginary time evolution using a semi-explicit backward Euler scheme. This scheme is efficient for the imaginary-time evolution of non-linear GPE-like equations~\cite {BaoJoSC2004,AntoineCPC2014}. After the change to imaginary time $t = i \tau$, the finite difference discretizations of the time derivative and application of the semi-explicit backward Euler scheme, we obtain an equation in the form
\begin{equation}
    \qty[\frac{\mathbb{1}}{\dd \tau} + \hat T + W(r) + G[f^n] \abs{f^n}^2] f^{n+1} = \frac{1}{\dd \tau} f^n,
\end{equation}
which is a linear matrix equation $\mathbb{A} \vb{v} = \vb{b}$, where $\mathbb{A}$ is a sparse matrix, $\vb{v}$, $\vb{b}$ are vectors; the superscript $n$ denotes the timestep. In each imaginary timestep $\dd \tau$, we solve the above equation to find $\vb{v}$. The sparsity of the matrix $\mathbb{A}$ allows us to employ efficient and robust Krylov solvers~\cite{Saad, Krylov}. 

\textit{Time evolution.} For dynamics calculations, we solve the real-time evolution of the system. For this purpose, the backward Euler scheme cannot be used as it is not energy-conserving. Note that this property of the Euler scheme is actually beneficial for ground-state calculation, as the additional damping of the total energy increases the speed of the convergence. However, to simulate closed-system dynamics, we need to ensure that the total energy is conserved. For this purpose, we employ the Spectral Time Splitting Method~\cite{BaoJoCP2006}.

Usually, in the Spectral Time Splitting Method, the kinetic operator is expressed in the momentum space using its Fourier transform. However, the standard form of this numerical scheme ensures that the wavefunction vanishes at the system's edges. It is not straightforward to introduce the boundary condition $\pdv{f}{r} \eval_{r=0} =0$ and $f(0) \neq 0$ in the Fourier space. Therefore, we treat the kinetic term in the position space as other operators. The main control parameter is energy error $\Delta E(t) = \abs{\frac{E_N(t)-E_N(0)}{E_N(0)}}$, where $E_N(0)$ is the total energy of the initial state with post-quench Hamiltonian, and $E_N(t)$ is the total energy at time $t$. We consider our results as stable as long as $\Delta E(t)<10^{-3}$. 

\begin{figure*}[t]
\includegraphics[width=0.96\textwidth]{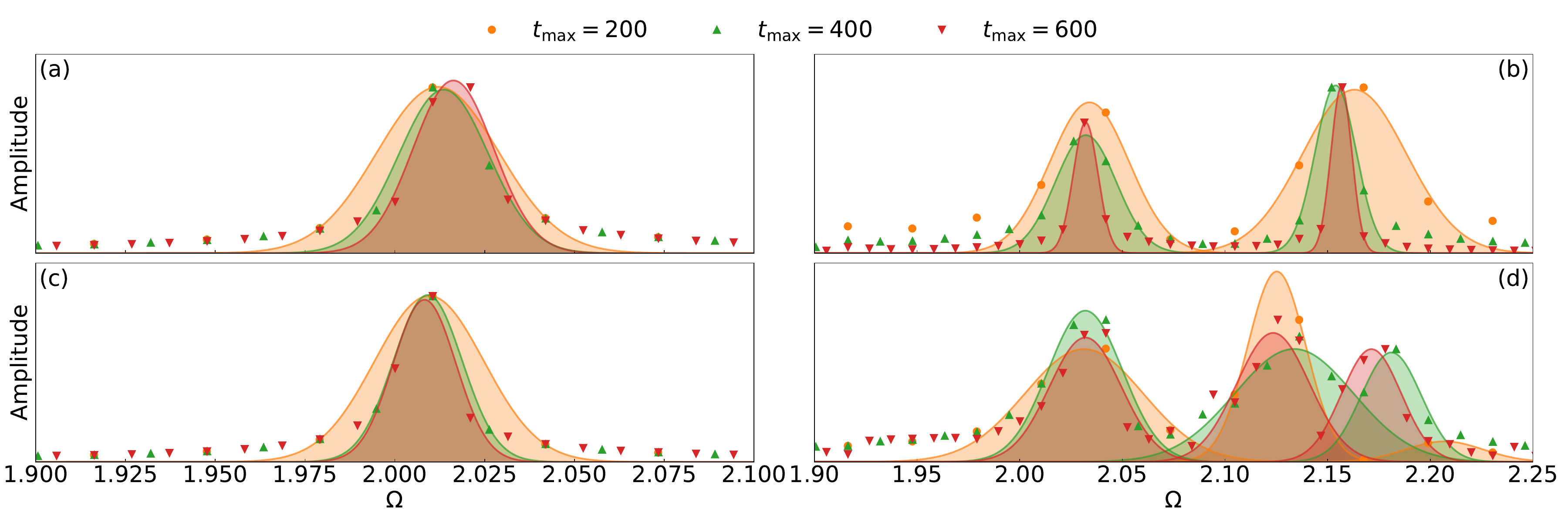} \\
   \caption{\textcolor{black}{The Fourier transform of $R^2$ signal (round and triangle markers) after the quench of the non-interacting state under interacting Hamiltonian with (a,c) $\ln B_2/N=-3.0$ (b,d) $\ln B_2/N=-2.14$ for number of bosons (a,b) $N=20$ (c,d) $N=50$ and for different propagation time from $t=0$ to $t_\text{max}$.  Solid lines with shaded area show a fitted Gaussian function to the peaks of the Fourier transform data.}  }
   \label{fig:sm_figurec}
\end{figure*}

As we cannot ensure stability in long-term propagation, we perform a Fourier transform of the resulting signal only up to the time point \textcolor{black}{$t_\text{max}=600$} that ensures that the energy error is below the chosen limit. \textcolor{black}{ We illustrate the dependence of our results on $t_\text{max}$ in Fig.~\ref{fig:sm_figurec}, which shows the Fourier spectrum and fitted Gaussian functions for different propagation times, $t_\text{max}\le600$. We observe overall agreement in both the number and position of the modes across the data for different values of  $t_\text{max}$. In general, we are able to perform calculations of the quenches from the non-interacting state to interaction strength $\ln B_2/N\le-2.14$ with reasonable stability. We can also study time dynamics at stronger interactions if the initial state has a higher overlap with the eigenstates of the post-quench Hamiltonian. For instance, we calculated the breathing dynamics (`perturbative quench') to determine breathing modes frequency.} 

Note  that if, for further investigation, longer stable propagation is needed, the energy conservation error can be reduced by improving numerical methods, for instance, by calculating the evolution of the equation
\begin{equation}
    i  \pdv{\psi}{t} = \qty[ -\frac{1}{2} \pdv[2]{\mathbf{x}}  + W(\mathbf{x})  +  G  N \abs{\psi}^2 - \mu(t)]\psi,
\end{equation}
where $\mu(t)$ is a chemical potential at time $t$. This scheme removes the trivial global phase associated with the stationary state~\cite{BaoJoSC2004, KoutentakisAtoms2022}.

\textit{Dynamic grid approach.} One of the main challenges in developing a robust computational scheme for our model is the presence of widely varying length scales in the solution, which depend on both the interaction regime and the number of bosons in the system. The size of the system $\ev{r^2}$ can differ by many orders of magnitude from $\ev{r^2} \simeq 1$ to even $10^{-20}$. To ensure that our program is robust and does not require users to estimate the system's size, we introduce a dynamic grid approach. Every $i$-th time step, the program estimates the size of the intermediate state $f^n$, adjusts the calculation grid so $\frac{\sqrt{\ev{r^2}}}{r_{N_\text{g}-1}} \simeq 0.03$, and extrapolates $f^n$ onto a new grid. We find that $i=5$ strikes a good balance between efficiently finding the proper grid size and keeping the overall computational cost low. This scheme ensures that we do not lose precision, that the state is well represented on the numerical grid, and that the system boundary at $r \to \infty$ does not introduce an artificial hard-wall potential. We continue calculations until the relative changes between time steps are $\Delta E$ and $\Delta \ev{r^2} <  10^{-8}$, where $\Delta O = \abs{\frac{O^{n-1}-O^{n}}{O^n}}$, which ensures the convergence to the ground-state. However, while this scheme ensures that results are not dependent on the initial guess of the system size, good initial conditions dramatically accelerate convergence. To use that, we calculate the ground states in the sequence. First, we calculate the ground-state for weak interaction $\ln B_2/N$, for which the convergence is quick, and then we use the solution and its size as an initial state to calculate the ground state for $\ln B_2/N + \Delta (\ln B_2/N)$. Repeating this procedure with gradually increasing interaction strength allows us to efficiently calculate the ground state over the full range of interaction strengths.


\textcolor{black}{Note that for real-time evolution, we do not employ the dynamic grid approach, as the extrapolation on the new grid introduces additional error, which would accumulate over the propagation time. This error is irrelevant to imaginary-time propagation, because it affects only the intermediate iterations, and we use the evolution to reach the converged ground state. Any interpolation error introduced during grid adaptation is removed as the system continues to evolve. }

\textcolor{black}{We find that in considered quench scenarios, the adaptive grid approach is not necessary.}   In the collapse dynamics, the total energy of the initial harmonic state usually prevents the system from collapsing to sizes that cannot be adequately described by the initial grid (which is chosen with respect to the harmonic state). However, this limits our ability to perform quenches to stronger interactions above $\ln B_2/N > -2.14$, where we observe that the calculations become unstable even on short timescales. This suggests a potential area for improving our numerical method to handle changing length scales better during real-time evolution. In breathing-mode calculations, we use the grid obtained from the ground-state calculations, which is already fitted to describe the system at the relevant length scale. This is because we expect the breathing mode to be a weak oscillation around the equilibrium state.  

\begin{figure}[t]
\includegraphics[width=0.48\textwidth]{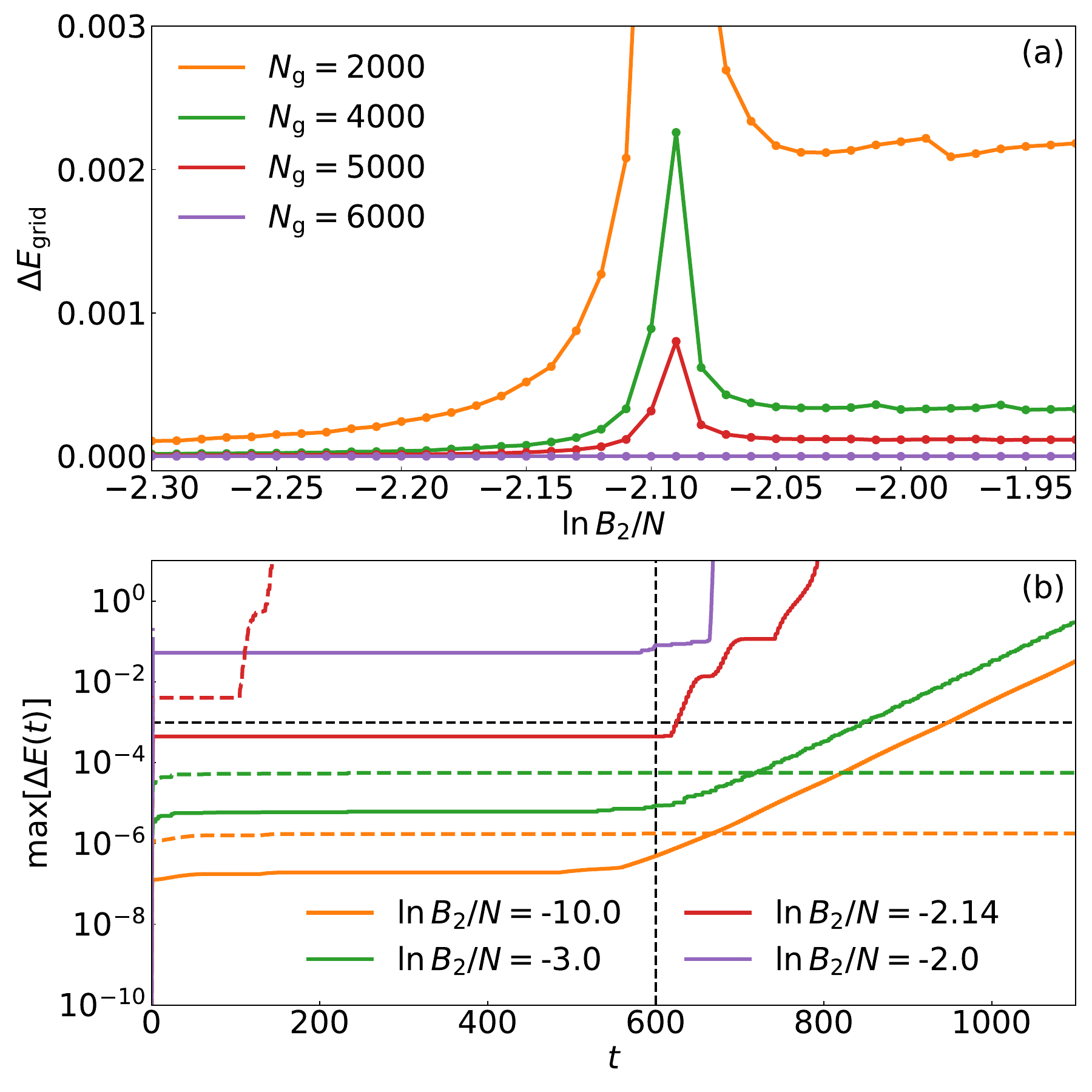} \\
   \caption{(a) Relative energy error $\Delta E_\text{grid} = |E(N_\text{g})-E(N_\text{g}=6000)|/|E(N_\text{g}=6000)|$ for the ground-state calculations with $N=50$ as a function of the interaction strength, $\ln B_2/N$. The peak at $\ln B_2/N \simeq -2.1$ is an artifact due to $E=0$ at this point. (b) Maximal energy error $\text{max}[\Delta E(t)]$ at time $t$ for quench dynamics with $N=50$. Solid lines show results for $N_\text{g}=6000$ and colored dashed lines for $N_\text{g}=2000$. The black vertical line shows the cutoff $t=600$, which we use for $N_\text{g}=6000$ to calculate the Fourier transform presented in Fig.~4 in the main text. Horizontal line $\text{max}[\Delta E(t)]=0.001$ shows the limit above which we consider our result unstable.  }
   \label{fig:sm_figure4}
\end{figure}

\begin{figure}[t]
\includegraphics[width=0.48\textwidth]{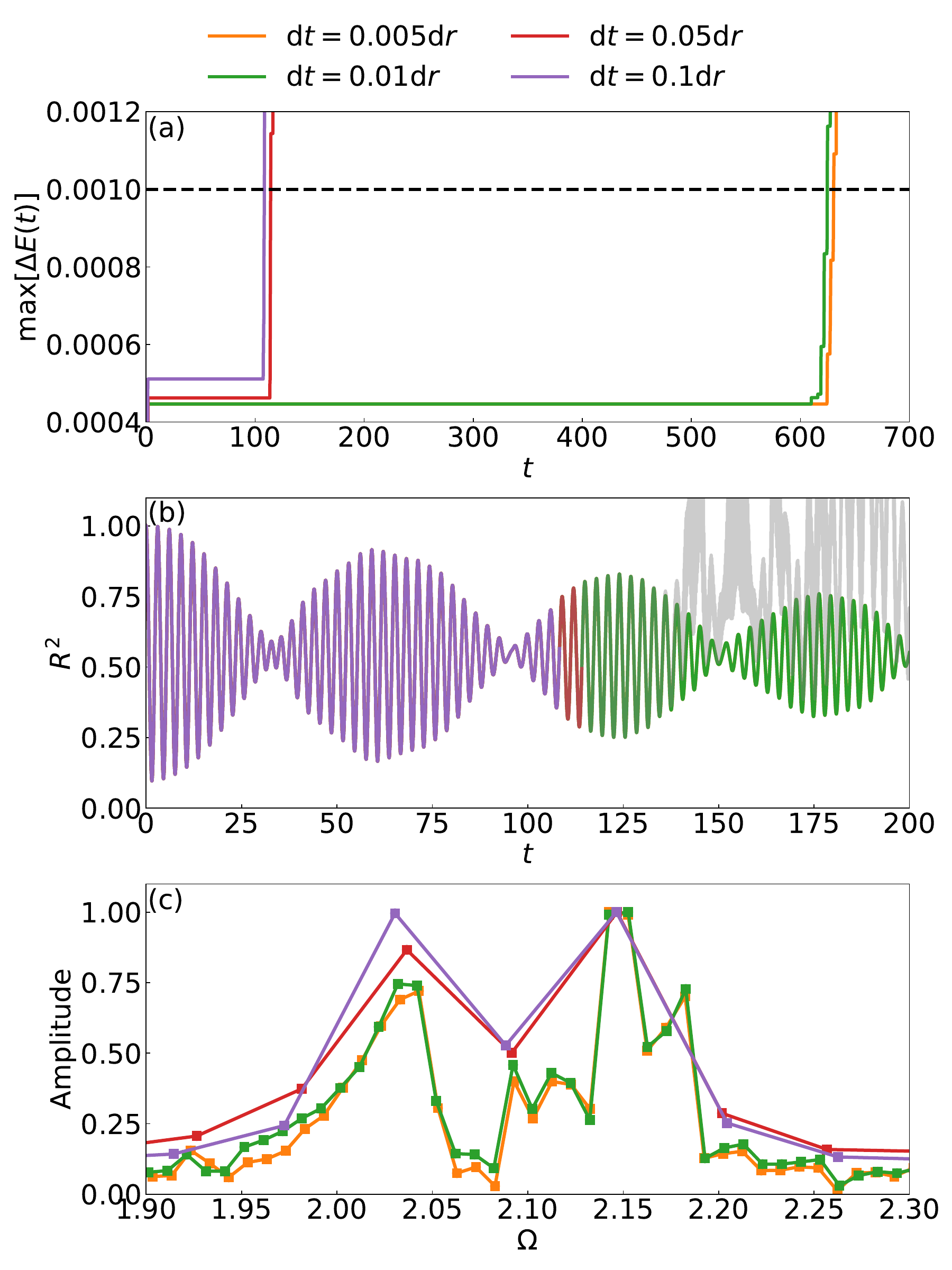} \\
   \caption{\textcolor{black}{(a) The maximal energy error $\text{max} [\Delta E(t)]$, (b) the size of the system $R^2$, (c) and its Fourier transform after the quench to interaction strength $\ln B_2/N=-2.14$ for $N=50$ for different time step $\dd t$ and grid size $N_\text{g}=6000$. In panel (b), solid colored lines show the evolution of $R^2$ until the time when $\text{max} [\Delta E(t)]>10^{-3}$. The gray curve show the evolution beyond this point in the numerically unstable regime for $\dd t=0.1 \dd r$. In panel (c), squared markers show Fourier transform data, while the solid lines guide the eye. }  }
   \label{fig:sm_figured}
\end{figure}

\textit{Computation parameters.} For both ground-state calculations and dynamics, we use $N_\text{g}=6000$, which gives optimal accuracy and efficiency of the calculations. Panel (a) in Fig.~\ref{fig:sm_figure4}, shows the relative error $\Delta E_\text{grid} = |E_N(N_\text{g})-E_N(N_\text{g}=6000)|/|E_N(N_\text{g}=6000)|$ for $N=50$ and for different interaction strengths $\ln B_2/N$. Note that at $\ln B_2/N \simeq -2.1$, the energy $E_N=0$, which causes the relative error to suddenly increase without really affecting the accuracy of the calculations.  For imaginary time evolution, we choose time step $\dd \tau =0.1 \dd r$ and for real time evolution $\dd t = 0.01 \dd r$. It is worth noting that decreasing the time step does not necessarily decrease accumulated energy error. For a time-dependent scheme, the grid size $N_\text{g}$ can directly and indirectly (through $\dd t$) affect the stability. We notice that for a smaller grid, the scheme is stable for a longer time $t$ because it requires fewer time steps, but only for weak interactions. For a larger grid, the overall energy error is lower, but the scheme becomes unstable in a shorter time $t$ because more time steps are performed, which accumulate errors. For stronger interactions, both stability and error are better for larger grids. The panel (b) in Fig.~\ref{fig:sm_figure4} presents the maximal energy error $\text{max} [\Delta E(t)]$ over time $t$ for $N_\text{g}=2000$ and $N_\text{g}=6000$ with $N=50$ \textcolor{black}{for the propagation of the non-interacting state under Hamiltonian with interatctions $\ln B_2/N$.} $\text{max} [\Delta E(t)]$ is defined as maximum value of $\Delta E(t)$ from the start of the propagation to time $t$. \textcolor{black}{Figure \ref{fig:sm_figured} shows the maximal energy error $\text{max} [\Delta E(t)]$, size of the system $R^2$ and its Fourier transform for $N_\text{g}=6000$ with $N=50$ for evolution of the non-interacting state to interactions $\ln B_2/N=-2.14$. We can see that the chosen $\dd t = 0.01$ is optimal, as further decreasing the time step does not yield a significant improvement in stability. Note that a smaller timestep does not increase the resolution of the Fourier spectrum, as it only depends on propagation time $t_\text{max}$. }

The code in the Julia programming language that allows for the reproduction of all our results is available as an open-source program on GitLab~\cite{Code}.

\section{Variational method}
\label{sec:townes}

\begin{figure}[t]
\includegraphics[width=0.48\textwidth]{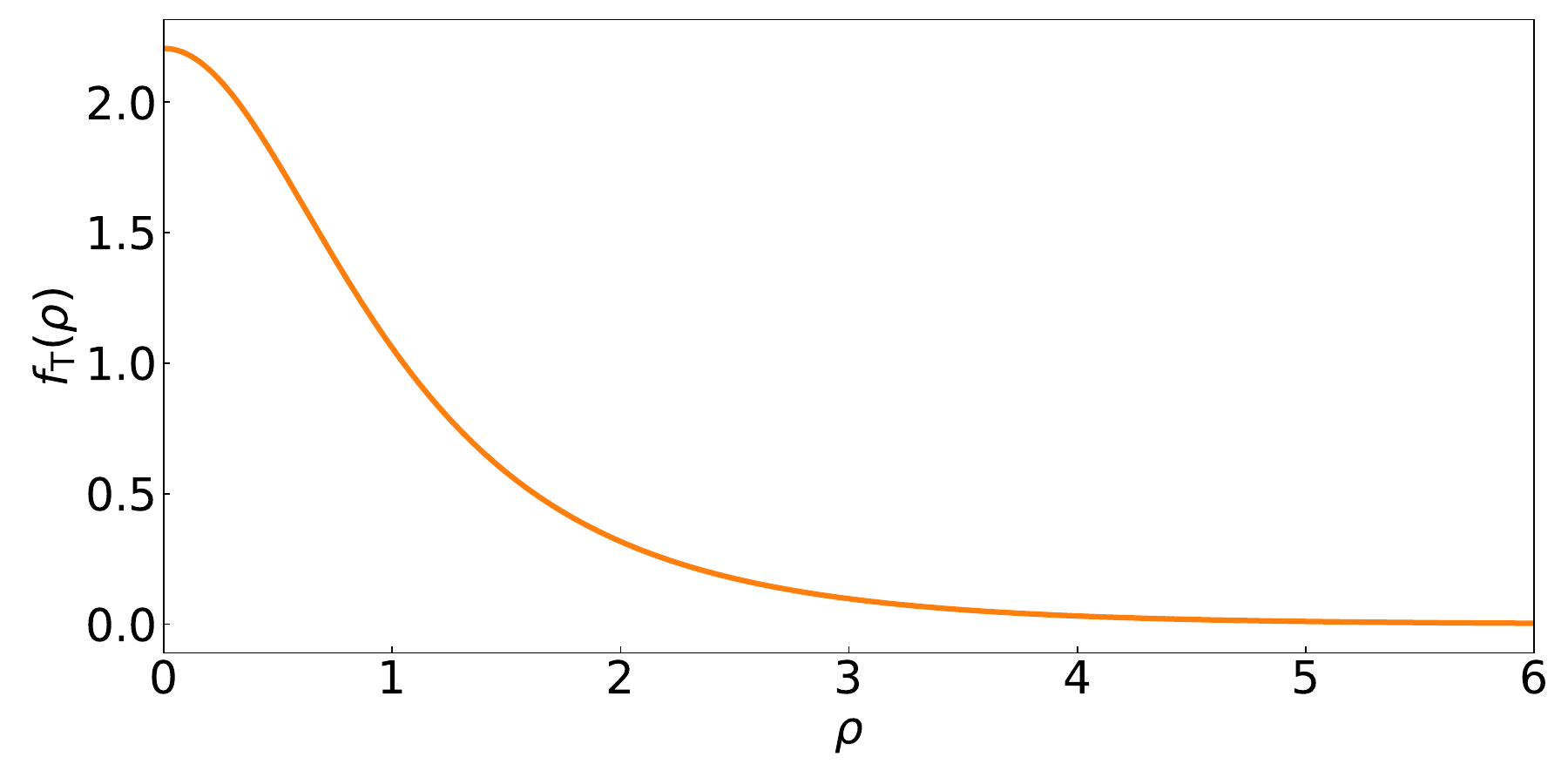} \\
   \caption{Function $f_\text{T}(\rho)$ determining the shape of the Townes soliton calculated as a solution of Eq.~\eqref{eq:sm_townes}.}
   \label{fig:sm_figure5}
\end{figure}

To compute the lowest-energy solution of Eq.~(3) from the main text variationally, we employ the scale-invariant Townes-soliton wavefunction~\cite{HammerPRL2004} $\sqrt{n(r)}=(2 \pi C \tilde R^2)^{-1/2} f_\text{T}(r/ \tilde R)$, where $C\simeq1.862$ and $f_\text{T}(\rho)$ is a universal function satisfying equation 
\begin{equation}
    \dv[2]{f_\text{T}}{\rho}  + \frac{1}{\rho} \dv{f_\text{T}}{\rho} - f_\text{T}(\rho) + f_\text{T}^3(\rho) = 0.
    \label{eq:sm_townes}
\end{equation}
This function has a characteristic bell shape (see Fig.~\ref{fig:sm_figure5}) with $f_\text{T}(0) \simeq 2.206$ and $\norm{f_\text{T}}=C$. The function $f_\text{T}(\rho)$ can be found by solving Eq.~\eqref{eq:sm_townes} using shooting method with boundary conditions $f_\text{T}(\infty)=0$ and $\dv{f_\text{T}}{\rho} \eval_{\rho=0} = 0$. This wavefunction has a single variational parameter, $\tilde R$, that determines the droplet's size; we numerically find the value of $\tilde R$ that minimizes the energy $E_N$ from Eq.~(1). Results of this variational treatment are shown in Figs.~2 and~\ref{fig:sm_figure6}.

The variational approach outlined here performs well across all interaction strengths because the Townes-soliton profile is exact in the strong-interaction limit and provides a reasonable description of the ground state in the weakly interacting regime. [A comparison between the shapes of Townes profile and the non-interacting harmonic oscillator wavefunction is shown in Fig.~2 of the main text.]

\section{Derivative of the energy}

\begin{figure}[t]
\includegraphics[width=0.48\textwidth]{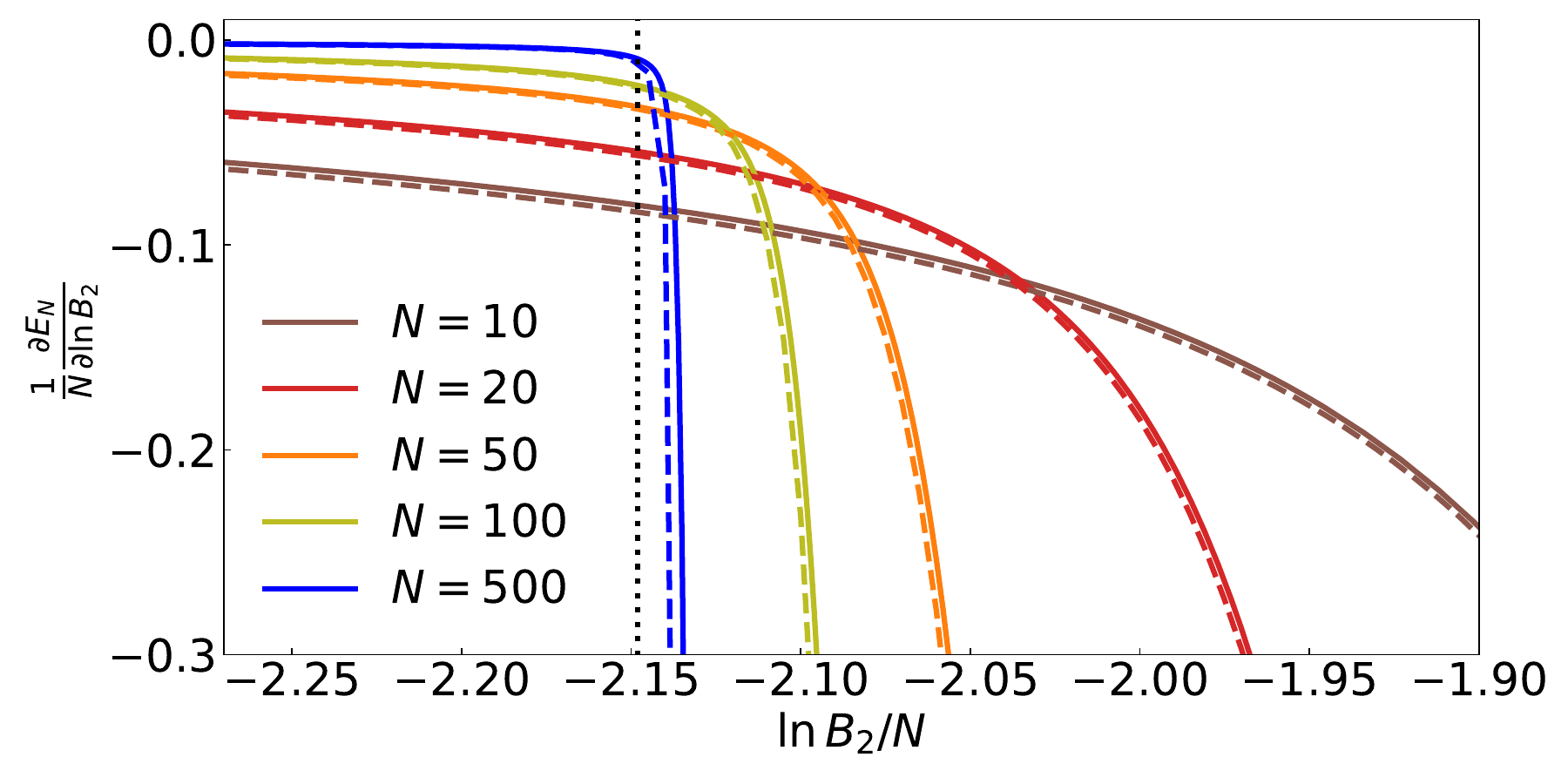} \\
   \caption{Derivative of energy over the interaction strength $\pdv{E_N}{\ln B_2}$ as a function of the interaction strength $\ln B_2/N$ for different number of bosons $N$. The solid curves show the exact numerical results; the dashed curves show the variational results with respect to the Townes-soliton profile. The vertical dotted line shows the transition point in the limit $N\to\infty$~\cite{Tononi2024,Brauneis2024}. }
   \label{fig:sm_figure6}
\end{figure}

Figure~\ref{fig:sm_figure6} shows the derivative of the energy $E_N$ over the interaction strength $\ln B_2/N$. The derivative is smooth across the transition; however, it is expected to diverge to $-\infty$ for $N \to \infty$ at the transition point $\ln B_2/N\simeq -2.15$. 

\section{Breathing dynamics}

\textcolor{black}{
\subsection{Calculations based on continuity equations.}} Similar to the standard mean-field description (see, e.g.,~\cite{Pitaevskii1996}),
the generalized Gross-Pitaevski equation leads to the continuity equation
\begin{equation}
\frac{\partial n}{\partial t}+\mathrm{div}\vb*{j}=0,
\end{equation}
where $\vb*{j}=-\frac{i}{2}(\psi^* \vb*{\nabla} \psi-\psi \vb*{\nabla} \psi^*)$, $n=|\psi|^2$,
 and the equation for the evolution of momentum
\begin{equation}
\begin{split}
\frac{\partial \vb*{j}}{\partial t} =  &-\frac{1}{2} \vb*{\nabla} \left(\psi^* T\psi+\psi T\psi^*+ 
\vb*{\nabla} \psi^* \vb*{\nabla} \psi\right) \\
  &-n \vb*{\nabla} \left[W+G N n \right],
\end{split}
\end{equation}
where $T$ is the kinetic-energy operator.

Let us calculate the time evolution of the size of the system, $R^2 = \langle r^2\rangle$. Using the continuity equation, we derive
\begin{equation}
\frac{\mathrm{d} R^2 }{\mathrm{d} t}=2 \int \mathbf{x} \vb*{j}\dd \mathbf{x}. 
\end{equation}
Differentiating this expression in time and using the equation for the evolution of momentum, we compute
\begin{equation}
\frac{\mathrm{d}^2 R^2 }{\mathrm{d} t^2}=4 \langle T\rangle - 2\langle \mathbf{x} \vb*{\nabla} \left[W+GN n \right]\rangle,
\end{equation}
where $ \langle T\rangle = E_\text{kin}/N$ is a kinetic energy per particle. For a harmonic oscillator potential, this equation can be further rewritten as
\begin{equation}
\frac{\mathrm{d}^2 \langle r^2\rangle }{\mathrm{d} t^2}=4 \frac{E_N}{N} - 4 R^2 - 2\langle \mathbf{x} \vb*{\nabla} GN n \rangle-2 \langle gN n \rangle.
\end{equation}
The last two terms do not cancel each other because of the broken scale invariance of the problem. This equation leads to Eq.~(13) from the End Matter ($N\gg 1$) 
\begin{equation}
\frac{\mathrm{d}^2 R^2 }{\mathrm{d} t^2}=4 \frac{E_N}{N} - 4  R^2 + 2 N  \ev{ n^2 \frac{\mathrm{d}g}{\mathrm{d}n} }.
\end{equation} 

 For the breathing dynamics, we assume that the solution has the form $\psi(r/R(t))/R(t)$. The time dynamics of $R(t)$ reads 
\begin{equation}
\ddot{R}^2 (t)=\frac{4E_N}{N} - 4  R^2(t) + \frac{N}{2\pi}{\langle n g^2\rangle}.
\end{equation}
For small oscillations $R=R_0+\delta R$, where $R_0=R(t=0)$ and $\delta R/R_0\ll1$, therefore, we can re-write this equation as
\begin{equation}
2R_0\delta \ddot{R} (t)\simeq -8R_0\delta R - \frac{N \delta R}{\pi R_0}{\langle n g^2\rangle}_{t=0}.
\end{equation}
The frequency of breathing oscillations is therefore
\begin{equation}
\Omega=\sqrt{4+\frac{N}{2\pi R_0^2}{\langle n g^2\rangle}_{t=0}}.
\label{eq:omega_SM}
\end{equation}
To arrive at the expression presented in the main text, note that $\langle n g^2\rangle_{t=0}=8\pi(R_0^2-E_N/N)/N$. 

\textcolor{black}{
\subsection{Calculations based on sum-rule approach.} There is an alternative sum-rule approach to determine the breathing mode frequency~\cite{BohigasPhysRep1979,dalfavo_RMP_1999,pitaevskii_book_BEC}. 
For an operator $F$, the strength distribution is
\begin{equation}
    S_F(\omega) = \sum_{n=1} \abs{\mel{n}{F}{0}}^2 \delta(\omega-\omega_{n0}),
\end{equation}
where $\omega_{n0}=\frac{E_{N,n}-E_{N}}{\hbar}$, and states $\ket{n}$ are eigenstates of the system with energies $E_{N,n}$ with ground-state $\ket{0}$ with energy $E_N$. The ratio of the weighted moments $m_{p+1}/m_p$ or $(m_{p+2}/m_p)^{1/2}$ corresponds to rigorous upper bounds for the frequency of the lowest monopole mode excited by the $F$ operator, where
\begin{equation}
    m_p = \int_0^\infty S_F(\omega)\omega^p \dd \omega.
\end{equation}
Moments $m_1$ and $m_3$ have known formulas valid for Hermitian operator $F=F^\dagger$ that can be determined using completeness of the eigenstates $\ket{n}$
\begin{equation}
    m_1 = \frac{1}{2} \mel{0}{\comm{F}{\comm{H}{F}}}{0},
\end{equation}
\begin{equation}
    m_3 = \frac{1}{2} \mel{0}{ \comm{\comm{F}{H}}{\comm{H}{\comm{H}{F}}} }{0},
\end{equation}
where $H$ is the Hamiltonian operator of the system.
Now, we consider $ F= F^\dagger= r^2$ so that $\ev{ F}=\ev{r^2}$. We look at the first commutator
\begin{equation}
\begin{split}
     \comm{ H}{ F} = \comm{ T}{ F} &=  -\frac{1}{2}(\vb*{\nabla}^2 r^2 - r^2 \vb*{\nabla}^2) \\
     &= -2 + 2 \vb{x} \cdot  \vb*{\nabla},
\end{split}
\end{equation}
which can also be expressed as
\begin{equation}
    \comm{ H}{  F} = -2i  D,
\end{equation}
where $ D = \frac{1}{2} \qty(\vb{x} \cdot \vb{p} + \vb{p} \cdot \vb{x} )$ is scale transformation (dilatation) operator~\cite{Pitaevskii1997,Camblong_JMP_2023}.
We find that the moment $m_1$ is 
\begin{equation}
    m_1 = \frac{1}{2} \ev{ F} = 2 \ev{r^2},
\end{equation}
because $\comm{ F}{\comm{ H}{ F}} =\comm{ F}{-2 i  D} = 4  F$ and $\comm{ F}{ D} = 2i  F$.
To determine the moment $m_3$, first we find that
\begin{equation}
    \comm{\comm{ F}{ H}}{\comm{ H}{\comm{ H}{ F}}} = 4 \comm{  D}{\comm{ H}{  D}}. 
\end{equation}
Now, we need to find $\comm{ H}{ D}$
with $ H=  T +  W + U$ with $W=\frac{1}{2}r^2=\frac{1}{2} F$ and $U=N\frac{g[n]n}{2}$. For operators $T$ and $W$, the commutators with the dilatation operator are known: $\comm{ T}{ D} = -2i  T$, and $\comm{ W}{ D} = 2i  W$.
To find commutator $\comm{D}{U}$, first, we consider the general case of the observable $A$ that transform as
\begin{equation}
    A \to A^\prime = U^\dagger(\theta) A U(\theta).
\end{equation}
For an infinitesimal transformation, we know that~\cite{Sakurai_Napolitano_2020}
\begin{equation}
    U^\dagger(\delta \theta) A U(\delta \theta) = A + i \delta \theta \comm{A}{ G},
\end{equation}
where $ G$ is a generator of the transformation. For the dilatation operator $ D$, this leads to the identity
\begin{equation}
    \dv{a} \ev{ A}_a \eval_{a \to 1} = i \ev{ \comm{ A}{ D}},
\end{equation}
where 
\begin{equation}
    \ev{ A}_a = \mel{\psi_a(r)}{ A}{\psi_a(r)},
\end{equation}
 is the expectation value of the operator $ A$ after scale transformation with factor $a$, where $\psi_a(r)=a\psi(ar)$. Note that commutators of $ T$ and $W$ agree with this identity as
\begin{equation}
    \pdv{\ev{T}_a}{a} \eval_{a \to 1} = 2a \ev{T} \eval_{a \to 1} = 2\ev{T},
\end{equation}
\begin{equation}
    \pdv{\ev{W}_a}{a} \eval_{a \to 1} = -2 a^{-3} \ev{W} \eval_{a \to 1} = -2 \ev{W}.
\end{equation}
For the operator $ U$ we find 
\begin{equation}
    \pdv{\ev{U}_a}{a} \eval_{a \to 1} = 2  \ev{U} + N \ev{n^2 \pdv{g}{n} }.
\end{equation}
which leads to 
\begin{equation}
    \comm{ U}{ D} = -2i  U - i N n^2 \pdv{g}{n} \equiv -2i  U - i  U^{(2)} .
\end{equation}
Then we find that
\begin{equation}
    \comm{ D}{\comm{ H}{ D}} = 4  H + 2  U^{(2)} +i \comm{U^{(2)}}{{D}},
\end{equation}
where 
\begin{equation}
\begin{split}
    i \ev{\comm{ U^{(2)}}{ D}} =& \dv{a} \ev{ U^{(2)}}_a \eval_{a \to 1} \\ =& 4 N \ev{n^2 \pdv{g}{n}} + 2 N \ev{n^3 \pdv[2]{g}{n}}.
\end{split}
\end{equation}
Finally, we see that
\begin{equation}
    m_3 =  8 E_N/N + 12 \ev{n^2 \pdv{g}{n}} + 4 \ev{n^3 \pdv[2]{g}{n}} ,
\end{equation}
where $\ev{H}=E_N/N$, which leads to breathing mode frequency
\begin{equation}
\begin{split}
    \Omega^2 &= \frac{m_3}{m_1} \\ 
    &= \frac{8 E_N + 12 N^2 \ev{n^2 \pdv{g}{n}} + 4 N^2 \ev{n^3 \pdv[2]{g}{n}}} {2 N\ev{r^2}}.
\end{split}
\end{equation}
We know that $E_N=NR_0^2-\frac{N^2}{2} \ev{n^2 \pdv{g}{n}}$ for time-independent problem with $\ev{r^2}=R^2_0$ and then we can transform
\begin{equation}
    \Omega^2 = 4 + \frac{N}{2 \pi R_0^2 } \ev{n g^2} + \frac{N}{4 \pi^2 R_0^2} \ev{n g^3} ,
\end{equation}
where $\pdv{g}{n} = \frac{g^2}{4 \pi n}$ and $\pdv[2]{g}{n} = - \frac{g^2}{4 \pi n^2}  + \frac{g^3}{8 \pi^2 n^2}$. If we approximate up to terms $\propto \mathcal{O}(g^2)$, we arrive at the same expression as Eq. \eqref{eq:omega_SM}
\begin{equation}
    \Omega  = \sqrt{ 4 + \frac{N}{2 \pi R_0^2 } \ev{n g^2} }.
\end{equation}
}

\end{document}